\documentstyle[here,12pt,epsfig,rotating]{article}
\input{psfig.sty}

\begin{document}

\vspace*{1.0cm}
\hspace*{10.8cm}VECC-NEX-98002

\bigskip
\bigskip

\begin{center}
{\Large {\bf A Preshower Photon Multiplicity 
Detector for the WA98 Experiment}} \\
\bigskip
\end{center}

M.M. Aggarwal$^{a}$, A. Agnihotri$^{b}$, Z. Ahammed$^{c}$,
P.V.K.S. Baba$^{d}$, S.K. Badyal$^{d}$, K.B. Bhalla$^{b}$, 
V.S. Bhatia$^{a}$, S. Chattopadhyay$^{c}$, A.C. Das$^{c,1}$,
M.R. Dutta Majumdar$^{c}$, M.S. Ganti$^{c}$, T.K. Ghosh$^{c,2}$, 
S.K. Gupta$^{b}$, H.H. Gutbrod$^{g}$, S. Kachroo$^{d}$,
B.W. Kolb$^{e}$, V. Kumar$^{b}$, 
I. Langbein$^{e,3}$, D.P. Mahapatra$^{f}$, 
G.C. Mishra$^{f}$, D.S. Mukhopadhyay$^{c}$,
B.K. Nandi$^{f}$, S.K. Nayak$^{f}$, T.K. Nayak$^{c}$,
M.L. Purschke$^{e,4}$, S. Raniwala$^{b}$, V.S. Ramamurthy$^{f,5}$, 
N.K. Rao$^{d}$, S.S. Sambyal$^{d}$, B.C. Sinha$^{c}$, 
M.D. Trivedi$^{c}$, J. Urbahn$^{e}$, Y.P. Viyogi$^{c,*}$

\smallskip

\begin{center}
\small\it{$^{a}$University of Panjab, Chandigarh 160014, India} \\
\small\it{$^{b}$University of Rajasthan, Jaipur 302004, Rajasthan, India} \\
\small\it{$^{c}$Variable Energy Cyclotron Centre, Calcutta 700064, India} \\
\small\it{$^{d}$University of Jammu, Jammu 180001, India} \\
\small\it{$^{e}$Gesellschaft f{\"u}r Schwerionenforschung (GSI), D-64220 
Darmstadt, Germany} \\
\small\it{$^{f}$Institute of Physics, Bhubaneswar 751005, India} \\
\small\it{$^{g}$SUBATECH, Ecole de Mines, F-44070 Nantes, France} \\
\end{center}

\bigskip

{\noindent $^*$ Corresponding author's address :} \\ 
{\noindent Variable Energy Cyclotron Centre \\
1/AF, Bidhan Nagar, Calcutta 700064 (India) \\
tel : +91 33 3371230, fax : +91 33 3346871 \\
e-mail : viyogi@vecdec.veccal.ernet.in}


\bigskip

{\noindent $^1$ Present address : Ohio State University, Columbus, Ohio, 
USA} \\
$^2$ Present address : KVI, Univ. of Groningen, Groningen, the Netherlands
\\
$^3$ Present address : IMK, FZK, D-76021 Karlsruhe, Germany \\
$^4$ Present Address : BNL, Upton, New York, USA \\
$^5$ Present Address : Department of Science and Technology, New Delhi
110016

\normalsize

\newpage

\begin{abstract}
     A high granularity preshower detector has been fabricated and
     installed in the WA98 Experiment at the CERN SPS for measuring the
     spatial distribution of
     photons produced in the forward region in lead ion induced
interactions. 
     Photons are counted by detecting the preshower signal in plastic
     scintillator pads placed behind a 3 radiation length thick lead
converter and applying a threshold on the scintillator signal to reject
the minimum ionizing particles. 
Techniques to improve the imaging of the fibre and
performance of the detector in the high multiplicity environment of
lead-lead collisions are described.
Using Monte-Carlo simulation methods and
test beam data of $\pi^-$ and $e^-$ at various energies the photon
counting efficiency is estimated to be 68\% for central and 73\% for
peripheral Pb+Pb collisions.

\end{abstract}



\bigskip

\section{Introduction}
      The primary goal of ultra-relativistic heavy ion collision (URHIC)
      studies is to investigate the properties of matter at high energy
      density and high temperature.
      These studies provide information on the 
      dynamics of multiparticle
      production mechanism and the phase transition from hadronic matter
      to Quark-Gluon Plasma (QGP). 
      Several hadronic and leptonic signatures have been proposed to
      search for the QGP phase transition.
      Because of their weakly interacting nature, photons are supposed to be
      the carriers of information about the initial
      state of matter\cite{dks}.
      Recently, URHIC experiments
       have drawn wide attention for the study of QCD vacuum
      through the search of disoriented chiral condensates (DCC). 
      Formation of DCC may lead to anomalous fluctuation of neutral 
      pion fraction \cite{anselm}, leading to large
      discrepancies in the relative  number of emitted charged particles
and
      photons. This is analogous to the Centauro and the anti-Centauro
      types of events observed in cosmic ray experiments \cite{centauro}.
The
      study of isospin fluctuations  entails the measurement of
multiplicities
      of charged particles and photons. 
      It is therefore important to study photon production in nuclear
collisions over
       a wide range of the phase space.

      A preferred technique for the detailed measurement of photons is to 
use   electromagnetic calorimeters, which are usually 15-20 radiation
lengths (X$_0$) deep \cite{saphir}. Here one measures the energies and
angles of emission of individual  
      photons.
However the large particle density in the
forward hemisphere 
      in URHIC experiments precludes the use of deep
       calorimeters because of enormous overlap of
      fully developed showers.
      The use of a preshower detector for measuring the multiplicity of
      photons in the forward hemisphere in  S$+$Au collisions  at the
SPS energy has already been
      demonstrated in the WA93 experiment \cite{wa93nim}. It was
      shown that using lead converter of thickness 3 X$_0$, the preshower 
photon multiplicity detector (PMD)
      can be designed to handle the particle density at a moderate distance of
      10 meters from the target, the resulting photon counting efficiency was
      $\sim$ 65\% for central collisions. 

      In the present article we describe the implementation of a preshower
PMD in the much higher particle density environment of Pb + Pb collisions
in the WA98 experiment \cite{wa98} at the SPS energy. 
The PMD employed plastic scintillator pads to detect the preshower signal
and wavelength shifting (WLS) plastic optical fibres to transport
scintillation light to the image intensifier (II) and CCD readout cameras. 
 It  was fabricated and installed in the experiment
in 1994. Simulation calculations for optimizing the phase space coverage
and  granularity of the detector are discussed in the next section. 
Results on  R\&D efforts to
      improve the performance  of the PMD are presented in
section 3. The
method of fabrication and
      quality control of the WA98 PMD is discussed in section 4. The
      performance of the detector in the test beam is described 
      in section 5. Parametrization of the PMD necessary for simulating
the effects of the entire detector are described in section 6. 
The method used for photon counting is described in section 7. 
Results on the estimation of the efficiency of
photon counting and the  purity of the photon
sample are also described here. A summary is given in the last section.

\section{ Simulation Results and Detector Layout}

	The main objective for the simulation studies in designing the
      preshower detector has been the optimization of converter thickness
and
      granularity to handle large particle densities. A major constraint
      was the limited number of readout cameras 
(image intensifier and CCD camera systems) available from the 
      UA2 experiment \cite{ua2}.
      The role of converter thickness was extensively studied during the
      design of WA93 PMD and the optimized thickness was found to be
3 X$_0$. Compared to S + Au collisions in the WA93 experiment, the
      particle density expected in Pb + Pb collisions at the SPS energy
was much
      higher. Hence a detailed simulation study was
      performed  to optimize the pad sizes and pseudorapidity coverage of
the detector using VENUS 4.12 event generator with
default parameter settings \cite{werner}
along with
the GEANT 3.21 detector simulation package \cite{geant}.

      Fig. \ref{density} shows the particle density (photons/cm$^2$) in
the forward
      hemisphere in central ($b<2$ fm) Pb + Pb collisions at 
      $158\cdot A$GeV
      as a function of  pseudorapidity ($\eta$)
      at  distances of 10 m and 20 m from the target
      as given by VENUS. The particle density in
      central $200\cdot A$GeV S + Au 
      collisions at 10 m from the target (corresponding to
      the location  of  the PMD in the WA93 experiment) is also
superimposed for       comparison. It
      is observed that (a) even at the lowest $\eta$- range under
      consideration, particle density in Pb+Pb collisions is comparable to
      that in the WA93 case only at a much larger distance from the
target,
      and (b) even at such a large distance the density is
      $2-3$ times higher at larger pseudorapidities, necessitating the use of
      smaller pads compared to those of the WA93 PMD in order to minimize
the    shower overlap.

      In addition to  the above requirements for handling  large
      particle densities, the detector layout and design was influenced by
 the following considerations :
(a) it  would be useful to
extend the detector
      coverage a little into the backward hemisphere ($\eta < 2.9$) to
      determine the peak of the pseudorapidity distribution properly,
      (b) the PMD with its
associated lead converter plates  and support structure should not spray
appreciable secondary particles onto the lead glass calorimeter sitting
just behind it;  (c) the PMD should have
full azimuthal coverage for the largest angular region possible; and (d)
the PMD should not receive appreciable spray of secondaries from the
upstream detectors and their support structures. 

      Studies of occupancy (ratio of the number of hit pads to the total
number of pads) 
and multiple-hit probability (as defined in \cite{wa93nim}) were carried
      out using VENUS and GEANT for square pads of  different 
sizes, considering only the PMD placed in air. The results are
      shown in Fig. \ref{multihit} for possible configurations using 15 
mm and 25 mm pads over the entire detector placed at 20 m from the target.
The results suggest that
      it was not possible to use the large pad sizes
      throughout the detector because of very high occupancy and
correspondingly large shower overlap. Using
smaller 
      pad sizes throughout was
      also ruled out because of the  requirement of very large number of 
      readout cameras. As the particle density varied by a factor of more
than 2 over the region of interest, it was considered appropriate to use
different pad sizes over different regions. Because of the large size of
the detector and the associated difficulties in handling the detector
during assembly, transport and installation, it was not
       convenient to have a single light-tight enclosure like that of the
WA93 PMD.  Therefore a modular approach was adopted, entire detector being
subdivided into box modules. Each box module was a
      light-tight enclosure, housing pads of a given size to be
readout using
      one II + CCD camera system. 
      It contained pads in a matrix of 38 rows and 50
columns. Thus the size of a box module was
determined by the size of the pads within it, varying from about 75 cm to 
125 cm in length and proportionately in width. This could be easily
handled during assembly and installation.

Several combinations of pad sizes and number of box modules of each kind
were tried keeping the phase space coverage and the total number of readout
units within the design goals.
    The final optimized layout of
the box modules  is shown in Fig. \ref{layout}. 
This contains square pads of sizes 15 mm, 20 mm and 25 mm, smaller pads
being placed in the innermost regions. 
The   box modules 21 and 22 as shown in Fig. \ref{layout} were made with
23 mm pads to cover
the gaps and achieve full azimuthal coverage without much loss of
      granularity. This resulted in a geometrically compact arrangement.
The PMD was finally located at 21.5 m from the target. 
The detector extends to approx. 21 m$^2$ in area and covers the
pseudorapidity
region 2.5 $\leq \eta \leq$ 4.2, of which
the region 3.2 $\leq \eta \leq$ 4.0 has full azimuthal  coverage.
With this arrangement for the entire PMD, consisting of 28 box modules,
the occupancy was  well
within 10\% and multiple hit probability within 5\%,
being  almost
      uniform over the entire  surface of the detector (see Fig.
\ref{multihit}).

  \section{ R\&D on Technology Improvement}

     The basic technique of fabrication and assembly of the PMD  in
various steps,
viz., cutting and drilling of scintillator pads,
     coupling the WLS fibre, assembly of pads in light--tight enclosures
and the use of perforated fibre end coupler (FEC) plate for arranging the
matrix of fibre bundles
 was similar to that used in the case of WA93 PMD
\cite{wa93nim}.
     Several improvements were incorporated at  various steps 
during the fabrication and assembly of pads
      so that attenuation of scintillation light passing through the fibre was
minimized and the
     image of the fibre on the CCD surface was focussed onto fewer number of
     pixels. R\&D results on these technological improvements 
 are described below.
Preliminary results have been
     reported earlier \cite{mrdm}.

     \subsection{ Techniques for minimizing  light attenuation}
\vskip 5mm

{\it     (i) Use of short length WLS fibre.}

     The emission peak of the WLS (BCF91A) fibre used in the present
detector lies at 480 nm. A
     substantial fraction of light around this wavelength is lost by
     self-absorption in the dye molecules of the WLS fibre after traversing a
     length of about 20 cm. Self absorption also results in shift of the
     mean wavelength to a larger value where the quantum efficiency of
the
S20 photocathode of the image intensifier
     is lower. Thus there is a substantial decrease in the detector signal
at
the end of the readout chain if long lengths of the WLS fibre are used, as
in the case of WA93 PMD.

     The length of the fibre required to
     transport light from the scintillator pads to the readout device in
the modular design discussed earlier
varies from 1.5 m for box
     modules containing 15 mm pads to 2.5 m for those containing 25 mm
pads. Such large lengths
are required to prevent the formation of
     loops  of smaller radii in the fibre, which might lead to loss of
light in the long run. 

While WLS fibre must be used inside the scintillator pad,
it is not  necessary to continue with the same fibre outside for
long lengths. In order to 
     improve  light transmission,  only
     8~cm long WLS fibres were used. These were joined  with clear plastic fibres.
 For mechanical protection and also to reduce the numerical aperture at
the exit end, clear fibres with
extra-mural absorber (EMA) coating were used. 
 As compared to the use of PVC sleeves to protect the WLS fibres,
as in the WA93 case, the EMA coating was preferred as the
overall thickness of the fibre remained small and managing the bundle
of 1900 fibres became relatively easy.
\vskip 5mm

     {\it (ii) Splicing of clear  fibre to the WLS fibre.}

The WLS fibre and the clear  fibre were joined using a specially
designed splicing jig shown in Fig. \ref{murthy2}. It     
 consists of two
     1 mm thick mating jaws made of copper, one movable and the 
     other fixed, attached with good thermal contact to a copper 
     block. The copper block is fitted with a 50 watt heater 
     cartridge and a temperature sensor. The jaws have small
     semi-circular cuts on the mating edges and in closed position
     form a circular opening of 1.5 mm diameter. The movable 
     jaw is spring loaded and remains in normally open position.
     It can be closed by a foot operated pedal arrangement.

\vskip 5mm

     The procedure for splicing is as follows : 

The ends of the clear and the WLS fibres to be joined are 
carefully cut normal to the
axis. EMA coating is removed from about 2 cm length of the clear fibre end.
     The prepared  ends of WLS and clear fibres are inserted
into
     a 1 cm long thin walled glass capillary tube of 1.02 mm inner
     diameter and 1.5 mm outer diameter. The 
     capillary tube is introduced into the cutout 
      of the heated (150  $^o$C) copper jaws (guided by a groove 
     on the teflon support platform of the jig). The jaws are closed 
     and the fibre junction is heated. After a 
pre-calibrated time (usually about a few
seconds)
     the jaws are released. The resulting spliced joint is  tested for 
mechanical strength by hanging a load of 500 g.
      The  joints are also subjected to light 
     leakage test by physically viewing the joint through a 20X
     microviewer. Bad joints which do not have 
sufficient strength or which show rings of
light around the splicing area are rejected. 
The short capillary tube is
retained over the splice and fixed with a drop of glue to improve the
strength of the spliced joint. 

     The fraction of light transmitted through the spliced joint 
(called  splicing efficiency in the present article) was
     measured using an LED/PIN diode setup 
similar to that used to characterize the fibres
in the case of WA93 PMD.  For this measurement a short (8 cm) piece of
clear fibre was spliced to another long (192 cm) piece of the same fibre.
In this system light was launched longitudinally 
into the short section 
 and the light output measured at the other end.
The result is shown in
     Fig. \ref{trans_coef} for a sample of spliced pieces. 
     Mean   splicing
efficiency  of more than  85\% was achieved even with such an
inexpensive jig and a simple splicing procedure. The width of the distribution,
which is a measure of the
piece--to--piece fluctuation in light 
transmission through the spliced joint, is found to
be less than 5\%.

For a quantitative estimate of the improvement of the detector signal
after
the use of splicing technique, the scintillator pad--fibre combination
was  irradiated with $^{106}$Ru
$\beta$--source and the detector signal measured using a photomultiplier tube
(PMT) setup. The mean signal for different sets of samples  are shown in
Fig. \ref{op_length} for the case of full length
WLS fibre (open circles) and the short  WLS fibre spliced with long  clear
fibre (open squares). Systematic errors, resulting from coupling of the fibre
to the PMT and the end finishing, are of the order of 10\%.
It is observed that
the signal decreases rapidly with increasing length for the full length
WLS fibre case. In contrast, the drop in signal with increasing fibre
length for the case of short length WLS fibre coupled
with long length clear  fibre is rather slow. In comparison to the full 
length WLS case, the signal for the spliced case increases marginally
for small total length and by more than
50\% for larger total length of the fibre, compensating for the sharp
drop in the earlier case.    
Thus the use of clear fibre makes the pad output almost
uniform for different  fibre sizes.

\vskip 5mm

     {\it (iii) Painting the tip of the fibre.}

     Light traversing within the WLS fibre towards the fibre 
end, which is inside the
     scintillator pad, emerges within the pad and is lost because of multiple
     reflection. A part of this light can be sent back into the fibre if the
     tip is shining and painted with scintillator grade reflector paint.

     The results on improvement in light collection arising from the use 
     of above procedures are also presented in Fig. \ref{op_length}. 
Improvement due to tip painting is 
found to be significant only when the total 
fibre length is small. For
larger lengths of the fibre, the improvement is only marginal. However for 
maintaining the
uniformity of  pad response for various fibre lengths,
      this procedure was introduced during the fabrication
     of the pads.

\subsection{ Improvement of fibre  imaging in readout}

	The image of the fibre on the CCD surface spreads over a set 
of pixels. The number of pixels per fibre is not uniform, being smaller
near the boundaries than in the middle.
Because of imperfections
in opto-electronic readout and  approximations used in the schemes for
generating pixel-to-fibre maps  \cite{wa93nim},
 some of the pixels belonging to a particular fibre may be assigned to
neighbouring fibres, giving signal in several fibres even though light
originated in only one pad.  
Thus light ``leaks out" to  neighbouring fibres. This 
effect has been discussed in
detail in the case of WA93 PMD \cite{wa93nim}. Another source of this light leakage
is the stray off-axis light in the fibre incident at the interface of the
entrance
window of the II chain. This off-axis light tends to broaden the image of
the fibre
at the CCD surface. To use the II + CCD device in the high
multiplicity environment of Pb + Pb collisions, 
the quality of the fibre image on
the CCD must be improved and the light leakage outside the assigned pixels
minimized.  Several techniques were investigated to reduce the
effect of off-axis light
in the fibre up to the surface of contact with the readout system.

Off-axis light traveling in the fibre is found to be substantially 
reduced by coating the
cladding surface of the fibre. This 
results in reduced numerical aperture. 
Light from the fibre enters the II
system through a glass window at the surface of contact of the FEC plate
and the II system. If  the fibre bundle is embedded in a  
transparent medium (as it was in the case of WA93 PMD), light reflected  
from the glass surface may
undergo multiple reflection and re-enter the II system, 
thus increasing cross-talk and broadening the image of the fibre. 
This will lead to further increase in shower overlap
in a high multiplicity environment. 
This effect  can be reduced by applying black paint over
the cladding surfaces near the exit end of the fibre bundle so that the
entire bundle becomes an opaque mass. The
results of the effects of applying
various paints to the fibres are presented in Table I in terms of the
effect on the measured numerical aperture.

The numerical aperture reduces slightly for larger length of the fibre 
 because of loss of some off-axis light at the
cladding--air interface over the length of the fibre.
This is evident by comparing the values of numerical 
aperture for 8 cm and 200 cm
long WLS  fibres. Application of paint to the fibre (on the cladding surface) 
also reduces the numerical aperture by cutting down off-axis light, 
black paint is more effective in this
regard. Splicing with clear coated fibre is much more effective in
reducing the numerical aperture at the exit end. With the removal of
off-axis light and
consequent reduction in numerical aperture the signal of the pad may also
decrease, but the image of the fibre on the CCD surface should be sharper.
Final results on improvement in imaging of the fibre on the CCD surface
are
discussed in sections 5.1.2 and 5.3.2.

\begin{table}[H]
\begin{center}
TABLE I \\
Numerical aperture of plastic optical fibres in various conditions. \\
\vspace{0.5cm}
\begin{tabular}{||c|c|c|c||}\hline \hline
 WLS fibre  &  Clear fibre & Fibre surface at & Numerical \\ 
 length (cm)  &  length (cm)            &  the exit end  & Aperture \\
\hline \hline
 8.0   &  -  & -- &  0.67      \\ 
 200.0  &  -  & -- &  0.60      \\ \hline 
 8.0   &  -  & White paint &  0.64  \\ 
 8.0   &  -  & Black paint &  0.62  \\ \hline 
 8.0  & 18.0  & Bare$^*$ &  0.61      \\ 
 8.0  & 36.0  & Bare$^*$ &  0.52      \\ 
 8.0  &  200.0 & Bare$^*$ &  0.50      \\ \hline \hline
\end{tabular}
\end{center}
$^*$ For these measurements,  ``bare" denotes the condition that EMA
coating was removed from about 5 mm
length of the clear fibre near the end. 
\end{table}

\section{WA98 Photon Multiplicity Detector}
\subsection{Mechanical description}

     \subsubsection{Fabrication of pads}

The  plastic scintillator sheets (3 mm thick BC-400) were
cut into square
pads of sizes 15 mm, 20 mm, 23 mm and 25 mm by using proper tools and
cooling procedures developed earlier during the fabrication of WA93 PMD.
Blind diagonal holes of 1.1 mm diameter were drilled into the pads. After
proper cleaning and drying of pads, a calibrated amount of optical glue
(BC600) was injected into the hole. Short 
(8 cm) pieces of WLS fibre,
with reflector paint  applied to one tip, were
inserted into the holes and lowered slowly till the end without trapping
air bubbles. After allowing for a curing time of about 24 hours,
the pads were painted with a thick coat of scintillator grade reflector
paint. Splicing of the clear coated fibres of desired length were then
carried out according to the procedure described in the previous section. Small
(10 cm ) pieces of black PVC sleeves were then fastened over the fibre
near the pad end to protect the WLS fibre section and the spliced joint
from mechanical damage and to prevent from picking up stray light. All these
operations
were carried out in a humidity and dust controlled environment.

Control on the splicing operation and check on reproducibility was carried
out by splicing samples of two clear fibre pieces and measuring the
splicing efficiency at regular intervals. 
For proper quality control and to maintain close uniformity among the
pads, sample detector pads were tested in well calibrated $\beta$--ray
test setup using $^{106}$Ru source. Suitable sampling was done in batches
for ensuring  proper quality  of pads of different sizes.  Sample
results of $\beta$-tests are
shown in Fig. \ref{beta} for 15 mm and 20 mm pads.
The x-axis denotes the 
 peak channel number of the digitized PMT signal.
The uniformity of the pads is better than 10\%. The response
of pads of different sizes is also very similar.
This makes the entire PMD 
uniform in response to minimum ionizing particles.

     \subsubsection{Assembly of detector box modules}

     A typical  detector box module, shown in Fig. \ref{module}, consists
of two parts : (1) a light tight pad enclosure of height 125 mm,
      made of 1.2 mm thick steel sheet, to house the scintillator pads
      and (2) a camera enclosure (700 mm length, 300 mm width, 300 mm 
      height) to house the II and CCD system. The dimensions of the pad
       enclosures differ according  to  the size of the  pads.
       The camera enclosure
was  identical for  all the modules. This was fixed  
      on  top of the pad enclosure with insulating studs for electrical
isolation. 
The top lid of the pad enclosure had a cutout for
the fibres to pass through. A light-tight rubber bellow, with an aluminum
collar and a fibre-end-coupler (FEC) plate, joined the pad enclosure to the
camera enclosure. The bellow assembly provided a flexible neck for the
bundle of 1900 fibres and formed an extension of the light tight enclosure
up to the interface of the II system.

A  brass sheet of thickness 0.1 mm, die pressed in 
the form of staircase steps, was attached to the bottom 
      face of the detector encloser to fix pads in rows and columns 
      leaving room to take out the fibre bundle above the pad planes.
Pads were fixed to the brass steps using colourless Silastic (RTV732)
sealant. Free ends of the fibres were taken through the bellows to the FEC
plate.
	The FEC plates having 1.1 mm diameter holes in a matrix of 38 rows and
50 columns were identical to those used in the WA93 PMD. Fibres were
inserted into these holes maintaining strict one-to-one correspondence
between the pad coordinates in the box and the coordinates of the
corresponding fibre on the FEC plate. 

The fibre coordinate at the top right corner of the FEC plate, as viewed
from the polished surface, is denoted (01,01). The corresponding position
on the pad enclosure was marked for proper installation and
identification with pixel coordinates on the CCD during generation of
pixel-to-fibre maps. 

After proper insertion of the fibres in the FEC plate, the bundle was
trimmed, potted, machined and then hand polished to a mirror finish to
achieve good light coupling with the glass window of the readout device.
For potting the fibre bundle a 3--component resin 
consisting of CIBA-GEIGY AY103 araldite, HY956 hardener and DW0137
black additive  was used. The black component
 made the  FEC plate opaque to light reflected from the glass window of
the readout device. This prevented 
multiple reflection of light and re-entry into the II system, thus
minimizing  cross-talk and broadening of the image of the fibre
on the CCD surface.
Finally the top lid
of the pad enclosure was fixed  in place with a black silicon sealant.
This  made the entire enclosure light-tight.

The front  window of the II + CCD camera was coupled to the
polished end of the fibre bundle
on the FEC plate and held in position with long screws, 
as shown  in Fig. \ref{photo1}(a).
A connector panel at the side of the camera enclosure, shown in  
 Fig. \ref{photo1}(b), was used to supply  voltages to the II chain
and the CCD electronics and to carry signals from the CCD. This also
provided feedthrus for
the optical fibre link from the xenon flasher to the local homogenizer
described below.

\subsubsection{Fiducial fibres}

In each box module there were 101 pads which had one additional
clear fibre inserted
for selective illumination. These pads, called {\it fiducial
pads}, were randomly but uniformly distributed over the matrix. The other
ends of the clear fibres of these fiducial pads (referred to as 
{\it fiducial fibres})  
       were bundled and potted with an optical glue
       into a small perspex disk of 20 mm
       diameter fitted with a brass collar and brought out through a 
       hole on the side of the rubber bellow. A 250 mm long and 20 mm
diameter 
      perspex homogenizer rod was coupled to the fiducial  fibre bundle. 
This system was
enclosed in a black plastic tubular housing to make it light tight. It was 
situated within the camera enclosure of each box module.
The other end of the homogenizer was connected to an optical fibre through
the connector panel in the camera enclosure.  This fibre 
formed one of the branches of a light distribution system and was
optically 
connected to the xenon flasher device.

Light flashes for the fiducial fibres in all the box modules of
 the PMD  were generated by an externally triggered flasher
 system as shown in Fig. \ref{xenon}. This consisted of a high efficiency
xenon lamp
(Hamamatsu L2445) having a stable arc size of 8.0 mm $\times$ 1.5 mm, a
perspex
homogenizer rod of 20 mm diameter and 300 mm length and a set of plastic
fibres (1 mm dia) to transmit light to various detector box
modules.

Measurement of the coordinates of the  fiducial fibres in terms of pixel
coordinates 
on the CCD surface was necessary to generate the
pixel-to-fibre map by the procedure described in Ref. \cite{wa93nim}.

\subsubsection{Support structure and detector installation}

     The  support structure for the PMD was made up of five sections of 
     6 mm thick steel plate welded together and  reinforced at the 
     boundary by rigid channels as shown in  Fig.
\ref{murthy3}(a). The
central section was perpendicular to the beam axis. The sections on the
two sides and on the top and  bottom were tilted inward by about 8$^o$
in order to maintain a near--normal incidence for all regions of the
detector.
     The entire support structure was fixed to a rigid stand
     fitted with rollers and could be moved longitudinally on
     rails to facilitate service access. 

Lead converter plates were mounted on the steel plate of the support
structure
facing the target. A set of fiducial markers were also fixed on the
support structure above the lead plates for proper geometrical alignment
of
the detector with
respect to the beam axis. 

Detector box modules were mounted on the back side of the support
structure in accordance with the layout scheme shown in Fig. \ref{layout}.
 Twelve box modules containing 15 mm pads, four box modules
containing 20 mm pads and two box modules containing 23 mm pads  were
mounted on the central
section of the support structure. The  side 
sections supported 
      modules containing 25 mm pads and  the top and 
     bottom sections supported   modules numbering 14, 15, 18 and 19.
     Fig. \ref{murthy3}(b) shows the
      detector box modules and the camera enclosures.
	In the box modules mounted on the top and bottom sections,
corners were cut and
sets of 12 $\times$ 12 pads
were removed  so that the PMD could be
moved very close to the lead glass calorimeter in the final position
without mechanical interference from its massive support structure.

One set of cables carried high voltage from the counting area to a
location near the detector. High voltage was then supplied to the 
individual
cameras through a set of distribution boxes. All the low-voltage supplies
for the CCD electronics were kept near the detector and their status was
monitored remotely. The length of the signal cables from the detector to
the digitizer was about 50 m.

The CCD electronics board of the II system was cooled by a built-in 
fan. In case of malfunction of this fan, camera noise would increase due
to rise in temperature. In order to monitor temperature variations of the
CCD
housing,  temperature sensors  (Analog Devices AD590) were
attached to the enclosure close to the CCD chip and the temperatures 
of all the camera enclosures were monitored via a remote digital readout.

  \subsection{Readout}

The II + CCD system of the UA2 experiment \cite{ua2} has been used for the
readout of the scintillator light signal as in the case of  WA93 PMD.
The 3-stage II chain provides light amplification of the order of 40000.
The CCD has a total of 31600 pixels arranged in rows, each row
having 218 pixels. 
A map assigning each pixel to a fibre, called the
pixel-to-fibre map, is generated to transform the CCD pixel coordinates
to the fibre coordinates of the PMD for clustering of preshower hits.

The CCD pixel charge is digitized using a custom-built FASTBUS module
using 8--bit 20 MHz flash ADCs.  
The reading of the FASTBUS module by the
data acquisition system and the corresponding timing diagram have been
discussed in detail in Ref. \cite{wa93nim}.
A free running clock of 1 $\mu$s width was used to
generate the clear signal in every 10 $\mu$s. This eliminated the CCD
noise stored during the intermediate time. The time difference between
the arrival of the clear clock and the event trigger was measured using a
TDC having a range of 10 $\mu$s. 
This helped in rejecting events which were partially
or fully cleared due to the arrival of the clear clock. Details of the
event clean--up procedure will be described in subsequent publications.

The digitizer takes data in three different modes \cite{wa93nim,ua2} :

{\noindent {\it Mode 1} : measures charges in all pixels without pedestal
subtraction. This mode is used to generate data on the pedestal charge
of the pixels and stored in the digitizer memory 
for use in Mode 2 and Mode 3.} \\
{\it Mode 2} : this gives digitized pixel charge after pedestal
subtraction.
Only pixel charges above a preset hardware threshold,
accompanied by the address, 
 are transmitted
during reading of the digitizer. \\
{\it Mode 3} : This gives fibre signal, i.e., sum of Mode 2 data for the
set of pixels assigned to a fibre as given by the pixel-to-fibre map.
Pixel-to-fibre map is loaded into the digitizer memory.
Fibre number and number of fired pixels are transmitted together
with the fibre sum.
This produces very compact
data format.

In addition to the
readout of the CCDs with the FASTBUS digitizers it was possible to view
the CCD images triggered by a timing generator in television mode (TV-mode).
This mode was used to inspect visually the camera gain and noise and
the alignment of pads to the test beam.

Small volumes of data, as generated during test beam and fiducial runs,
were collected in Mode 2. In the main experiment the data volume of the
PMD was large and hence it was necessary to use Mode 3 option only.
The pixel-to-fibre maps were generated using fiducial data
taken at regular intervals and loaded into the digitizer memory.

\section{Detector Characteristics and Test Beam Results}

         The characteristics of the PMD, e.g., response to varying
    input light levels, minimum ionizing signal and preshower
    behaviour, were studied using test beams and Monte-Carlo
    simulations
    of known particle types ($\pi^-$ and $e^-$) with different
    incident
    energies and also using fiducial lighting at various light
intensities. 
    The results will be described by using the terms  ``ADC", ``pad ADC''
and
    ``cluster ADC'' \cite{wa93nim}.  ADC is used here to denote the ADC
channel number of
the digitized data. Pad ADC denotes 
    the sum of the ADC channel numbers of the pixels (digitized
charge after pedestal subtraction) assigned to a particular 
    pad (fibre), the term ``cluster ADC'' denotes the sum of the 
    ADC channel numbers of all the pixels in a cluster in mode 2 or the
sum of 
    pad ADCs of all the pads in the cluster in mode 3. 
    The cluster
    may be that due to a hadron (and its secondaries) or  a photon
    preshower or  a lighted fiducial fibre.

\subsection{ Results from fiducial lighting}

Studies with fiducial lighting were carried out for all the
box modules fitted with readout cameras.
The results are described below.

\subsubsection{Study of camera noise}

         Using data on fiducial lighting
      the noise behaviour of the II + CCD  readout
cameras has been studied.
      In the  fiducial pattern  a control zone is selected in which a
fibre 
      is isolated from the neighbouring fiducial fibres 
      at least by $\pm2$
      pads. In principle no signal is expected in this 
      zone. These zones were studied very carefully for all the cameras. 
It was  found
      that these zones contain mostly single pixel clusters with $\le$ 4
ADC channels, as was also found in the case of WA93 PMD.
Mean ADC of the clusters in different cameras varied from 3.5 to 4.1. 
       This noise is rejected in the offline
      analysis before processing the data for physics studies.

\subsubsection{Light leakage}

         The finite resolution of the opto-electronic readout causes a
distortion in the image of the rectangular pattern such that the
assignment of pixels to a fibre is non-uniform, the number of pixels per
fibre being lesser at the boundaries than in the central region of the CCD
surface. In addition off-axis light in the fibre causes spreading of the
image. 
This results in
      a fraction of the light from a particular fibre to be detected
outside the set of pixels 
      assigned to that fibre. The neighbouring fibres thus appear to give
signal even though the corresponding pads have no light. This leads to
artificial spreading of the image of a fibre due to the readout system. 
The fraction of light  ``leaking
out" also depends on the intensity of input
light \cite{wa93nim}.
       It is therefore necessary to study the behaviour of
      light leakage to the surrounding fibres.

 The leakage was studied by considering a set of doubly isolated 
      fiducial fibres in fiducial data, there being only one fiducial
fibre in a 5 $\times$ 5 matrix. Thus the effect of light leaking from the
central fibre to the surroundings in a 3 $\times$ 3 matrix can be studied,
as these fibres are not expected to be affected by light from other
fiducial fibres.
The input light level was varied by adjusting the
intensity of the flasher system. The light leakage as a function of the
intensity (represented by total
ADC of the  cluster) 
       was found  to be more in the readout 
      direction at all light levels, as reported earlier \cite{ua2}. 
      The percentage of  leakage (total ADC channel numbers in the
surrounding fibres with respect to  the cluster ADC), shown in
Fig. \ref{adcleak},
slowly increases 
      and  then saturates with increase in  input light intensity.
Minimum leakage at low intensity is $\sim$25\% and
the saturation value is $\sim$40\%.

         It is also observed that the light leakage for the WA98 PMD is
less than that of the WA93 PMD at all light levels.  The saturation value
is 
about 
      30\% less compared to the WA93 case and occurs at much  higher
light levels (beyond
500 ADC channels). This
demonstrates the improvements in the detector as described in section~3.

\subsubsection{Spatial spreading of fibre images}

Light leakage discussed above causes the image of  a fibre to spread to
several neighbouring fibres. The number of extra fibres giving signal
depends on the intensity of incident light. This has been studied in
detail for several doubly isolated fiducial fibres. The mean probability
of a number of extra fibres receiving leaked light is plotted in Fig.
\ref{spatial} for
a set of light intensities, denoted by the cluster ADC values.

At lower light levels an average of 4 to 5 extra fibres
are affected by the leaked light. The distribution is almost symmetric. As
the light intensity increases,
 the maximum of the  probability distribution shifts towards larger
number of surrounding fibres and the distribution becomes asymmetric.
Beyond a
certain light intensity all the eight surrounding fibres
receive leaked light.

Because of unequal distance of  perpendicular and diagonal fibres in a
rectangular matrix and also because of the effect of higher charge
collection along the readout direction, fractional distribution of leaked
light into the surrounding fibres is not uniform. The distribution of
leaked light into various neighbouring fibres has  been
studied  using fiducial data at several light intensities.

\begin{table}[H]
\begin{center}
TABLE  II\\
Percentage distribution of leaked light to various fibres.
\end{center}
\end{table}
\setcounter{figure} {-1}
\begin{figure}[H]
\vspace*{-1cm}
\centerline{\epsfig{file=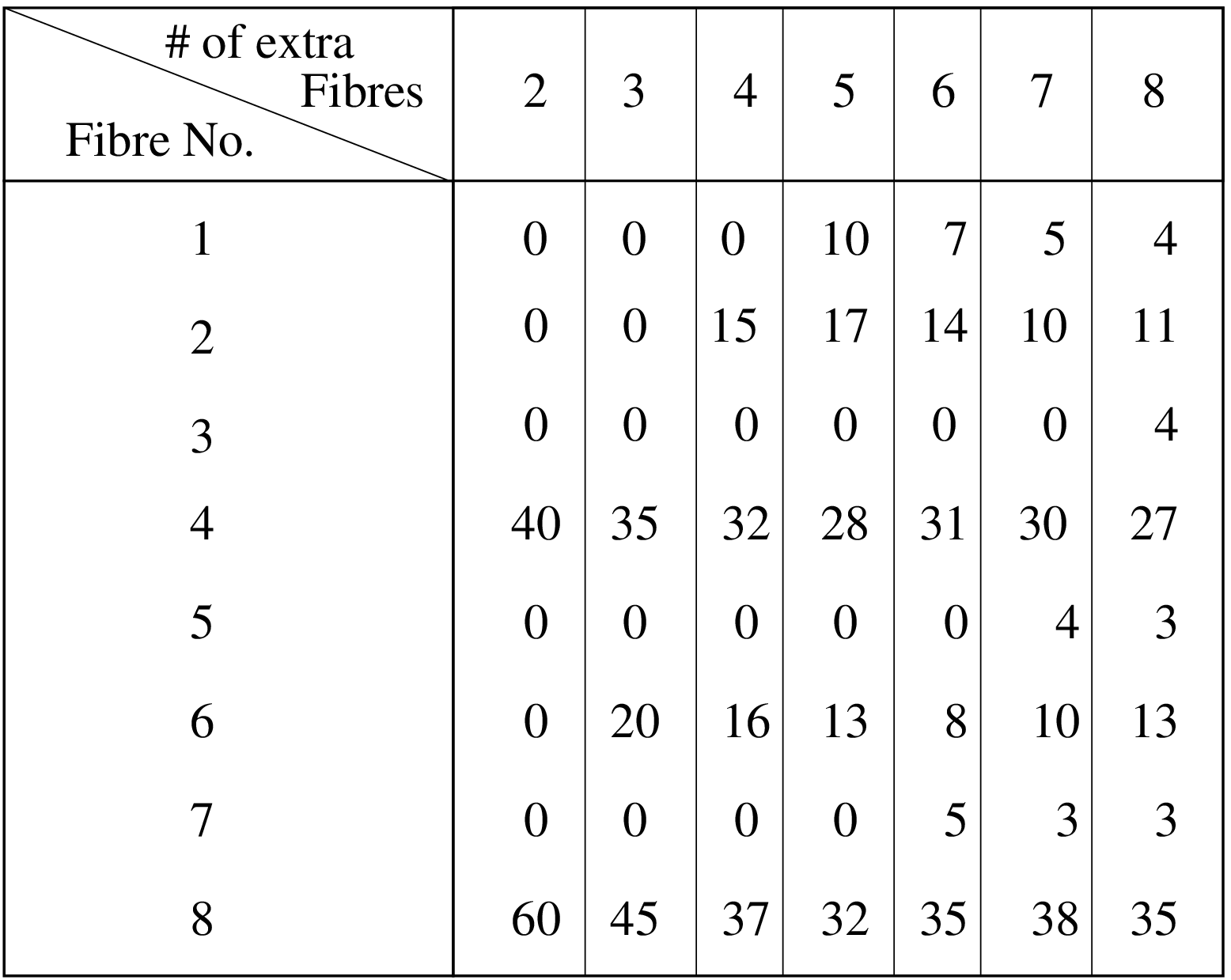,width=8cm}}
\end{figure}
\setcounter{figure} {0}

Fig. \ref{pattern} displays the nomenclature for addressing the fibres
around a fiducial one in a 3 $\times$ 3 matrix.
Table II summarizes the results on the percentage of leaked light going
into a
given fibre for various cases of the number of extra fibres affected. 
It is found from  Table II that fibres situated at the corners of
the rectangular matrix receive least light. Among the other four fibres
those
in the horizontal row receive more light than the fibres vertically
opposite the fiducial  one. The fibre situated to the right of the
fiducial (\# 8)  receives maximum light in all the cases.

\subsection{Test beam setup }

The tests of the PMD were carried out in the X1 beam line at the CERN SPS
using secondary beams of $e^-$ and $\pi^-$ of varying  energies in the
range $3-50$ GeV. Due to the problem of enormous time required in
exposing all the pads individually to the beams, only one box module of 15
mm pads was tested in the beam. Information on the behaviour of all the 
other
box modules was obtained by comparing the results of fiducial lighting
and
Monte-Carlo simulations with those for the tested box module.
         
The box module was positioned vertically on a remote controlled movable
stand in the X1 beam line.
The initial position of the beam was fixed on a 
      particular fiducial pad by viewing  the image of the fibre in TV
mode.
      The stand was then moved in suitable directions to bring other
pads into the beam for tests.

      Triggering arrangement consisted of a set of three scintillator 
      paddles. One scintillator paddle was placed in front of the 
      beam exit and the other one  at the detector box. The third 
      one with a hole was placed in between these paddles. For a 
      valid trigger, the two outer scintillators were required to be in
coincidence between them and in
       anti-coincidence with the paddle having a hole.

          Signals from Nitrogen and Helium filled Cerenkov counters 
      placed upstream and a delay-line wire chamber placed in front of the
detector were recorded for
offline use.
Cerenkov signals were used  for the separation of
electron and
pion beams. The triggering allowed a small ($<$ 1\%) admixture of
electrons
in the pion sample.
          The beam profile was obtained from the wire 
      chamber.
     The  offline data analysis selected only those hits of 
 the beam particles  which were located
       in the central position of the wire chamber profile. 

\subsection{Test results}

    \subsubsection{Minimum Ionizing Particle (MIP) signal}

Most of the charged hadrons impinging on the PMD deposit energy equivalent
to a minimum ionizing particle (MIP).
A detailed knowledge of the MIP response on the PMD is necessary for
parametrization of the detector simulation, for
rejection
of hadrons by applying a threshold on the cluster signal and for the
estimation of photon counting efficiency \cite{wa93nim}.

          The response of the pads to minimum ionizing particles
      was studied using 50 GeV $\pi^-$ beam without lead converter. 
      All the MIP properties like the number of fibres affected and the 
      percentage of light leaking to the surrounding fibres were studied 
      using the $\pi^-$ test data. About 30\% of the MIP clusters are
confined
to one fibre (pad) only.
      A typical MIP spectrum is shown in Fig. \ref{mipspectra}.
	Describing the MIP spectrum by a Landau distribution, the mean is
found to be $39\pm 1$ ADC channels.

           The single photoelectron response of the II + CCD system is
studied by fitting
       multiple Poisson  distributions  to the MIP spectrum.
      The mean value of the first
        Poisson component (with n=1) is found to be  $12.2\pm1$ ADC
channels, which
represents the signal due to 
       a single photoelectron produced at the first photocathode of II 
chain. This value is very close to that observed earlier in the case of
the
WA93 PMD, suggesting that the camera gains were very similar in the two
cases.
Considering the above value of
      the single photoelectron response, MIP signal is equivalent to  
      $\sim$ 3.2 photoelectrons. 
The number of photoelectrons in the
$\beta$-spectrum of  scintillator pads, measured during fabrication, was
found to be $\sim$4 \cite{mrdm}. The above two
results are consistent with each other considering the fact
that (a) some reduction in light output would have resulted in the
assembled
version of the detector because of the black paint, and (b) the energy
deposition in
$\beta$-source tests  is not truly MIP-like, but slightly higher, because
of contributions from
 very low energy $\gamma$-rays in the radioactive source.    

The MIP response suggests that the light output of the WA98 PMD is 
higher than that of the WA93 PMD \cite{wa93nim}. However,
the effect is not as
large as would be suggested from Fig. \ref{op_length}. The reason can
again be traced to the reduction in off-axis light by the  paints on the
fibre surface and at the ends.
Considering the requirement  of dynamic range, as discussed in
\cite{wa93nim}, the present value of the MIP signal is quite satisfactory,
a large increase would have resulted in saturation of the pad signal in
many cases.

    \subsubsection{Preshower characteristics}

          Preshower characteristics of the PMD, like the energy deposition
and transverse spread (number of pads fired),
       have been studied using a 3 X$_0$ thick lead converter placed in
front of the detector box module bombarded with 
       electrons of  varying energies in the range 
       3 GeV to 50 GeV. 

	Data analysis was carried out after noise filtration in mode 2 
data and
transformation to mode 3 (fibre or pad coordinates) by 
      using the appropriate pixel-to-fibre map generated from the 
      fiducial data taken during that period. A zone of 
      $10\times10$ pads around the centroid of the beam was chosen in the
      pad space. The pad hits were analyzed in two  ways to
      study the systematic uncertainty in the mean number of pads
      fired and the mean energy deposition for different electron
energies.
       In the first method total cluster signal was
obtained simply  by adding the signals in hit pads in the  
      selected zone. In  the second
      method the cluster signal was obtained by using proper clustering
algorithm developed earlier \cite{wa93nim}.
The cluster signal obtained from the two methods differ by less than 5\%,
the clustering method yielding a slightly lower value. 
The average values of
the two methods have been taken for the analysis.

          The detector simulation has been carried out using GEANT 
      for single particle $\pi^-$ and $e^-$ at specified energies.
      The systematic uncertainty in the mean energy deposition 
      due to the variation in  cut-off energies of the
      secondaries and
      $\delta$-rays produced in the lead converter 
      has been studied in detail.   The cut-off parameters 
      have been optimized to values which do not affect the  
results on energy deposition
and also minimize computation time. 
Discussions on this are presented below.

\vskip 5mm

{\it (a) Transverse shower spread}

The number of pads hit is a measure of the transverse shower spread.
 Fig. \ref{padsimdata}(a) shows the variation of mean number of pads (of
size
15 mm) for
different electron energies. The GEANT simulation results are also
superimposed for comparison. 
          It is found       that the total number of pads
      fired in the test beam is more than that fired
      in  the simulation.
      This is mainly because of  light leakage arising from 
      imperfections in pixel-to-fibre maps and the distortion of the
image on the CCD surface
      as discussed earlier. 

      It is observed that the mean number of pads fired 
is much less compared to those in  the    WA93 PMD
for all     electron energies, 
even though the pad area in the present case is almost
half of that 
 in the WA93 case. A comparison of the GEANT simulation and test
beam results of WA93 and WA98 PMDs is presented in Table III. It is
observed that  while the number of pads fired in
test beam data 
 at any particular electron energy 
was more than four times that of GEANT simulation in the WA93
case,  in the present case the number of pads increases only by a factor
of about two in going from GEANT to test beam results.
The transverse spread of the preshower is
greatly reduced because of the technological  improvements in fibre
imaging. It is to be noted that in the case of the WA93 PMD the potting
of the FEC
plate having the fibre bundle was done using a transparent resin, while in
present case it was black resin.

\begin{table}
\begin{center}
TABLE III \\

Comparison of average number of pads fired in GEANT simulation and test
data for various electron energies.
\vskip 5mm
\begin{tabular}{|c|c|c|c|c|}    \hline
Energy                 & \multicolumn{2}{c|}{WA93 PMD} 
                 & \multicolumn{2}{c|}{WA98 PMD} \\
(GeV)                & \multicolumn{2}{c|}{(20 mm pads)} 
                 & \multicolumn{2}{c|}{(15 mm pads)} \\
\cline{2-3} \cline{4-5}
                 & \multicolumn{1}{c|}{Simulation} 
                 & \multicolumn{1}{c|}{Test beam} 
                 & \multicolumn{1}{c|}{Simulation} 
                 & \multicolumn{1}{c|}{Test beam}         \\ \hline
3 & 3 & 16 & 4 & 7 \\
5 & 4 & 18 & 5 & 9 \\
10 & 4.5 & 19 & 6 & 13 \\
20 & 5 & 21 & 7 & 15 \\ \hline
\end{tabular}
\end{center}
\end{table}

{\it (b) Energy deposition}

The mean cluster ADC values are plotted as a function of incident electron
energy
in Fig. \ref{padsimdata}(b).  The  average energy deposition increases
linearly
only up to about 10 GeV and then starts becoming non-linear as also
observed in the case of WA93 PMD. 
This non-linearity arises because of two reasons : (a) the number of
shower particles does not increase linearly with  energy  at
higher incident energies, and (b)  light emission in the scintillation
process
is known to be non-linear when the energy deposition is high.
For the present applications,
however, this is not a significant problem as the cluster signal is used
only for rejecting hadrons by applying a suitable threshold. This
threshold, to be discussed in a later section, is rather small,  not
affected by non-linearity occurring at high signal levels.   

          The relation between the energy
deposition in a pad in MeV units as given by GEANT and the ADC channel
number of the CCD charge is displayed in  Fig. \ref{mev-adc}.
      The lowest point corresponds to the energy deposition by a MIP 
(50 GeV $\pi^-$ beam)  and the other points correspond to electrons
      at different incident energies. A   second order polynomial 
       is fitted through the points to get the MeV-ADC
      calibration relation. This relation is used to convert
      the GEANT energy deposition per pad in MeV  to  ADC channels in
order to generate simulated PMD data as described later.
The uncertainty in the MeV$\rightarrow$ADC conversion is less than 5\%.

\subsection{Comparison of test data and simulation}

\subsubsection{Readout resolution}

           After converting the energy deposition in MeV obtained from 
      simulation to  ADC channels 
       the simulation energy spectra  are compared 
      with the test beam energy spectra for various electron energies as
shown in 
      Fig. \ref{adcspectra}. The dashed histograms represent 
      the simulation energy spectra (in ADC units) and the 
      filled circles are the test beam data. It is observed that 
       the width of the test beam  data is
      larger compared to the width of the simulation energy spectra  in
all
the cases. 
      The increased width of the energy spectra in the test beam data is
because of
      the extra fluctuations in the production of scintillation
light, collection and transmission of light
      through the WLS and clear fibres and in various stages of
propagation of light 
      through the II chain to the CCD. This is in addition to the
fluctuations in
      the energy deposition in the sensitive medium, 
which is well described by the GEANT simulation.
      For the present discussions  all other contributions are termed  
      as  {\it readout resolution}. 
      Assuming that the various contributions to the fluctuation are
uncorrelated and the
widths  add up in quadrature,
one can deduce the magnitude of contribution due to readout resolution
using the relation

$       {\rm Readout~Resolution} = 
       \sqrt{(\frac {\sigma_{\it test}}{\Delta E_{\it test}})^2 - 
       (\frac {\sigma_{\it sim}}{\Delta E_{\it sim}})^2} $

{\noindent where $\Delta E_{\it test}$ and $\Delta E_{\it sim}$ are the
mean energy deposition obtained from test results and simulation (after
applying MeV-ADC conversion) respectively; $\sigma _{\it test}$ and
$\sigma _{\it sim}$ are the widths of the distributions in the two cases.

      The readout resolution as a function of cluster ADC as obtained 
      from the test beam data is shown in 
      Fig. \ref{readoutres}. Only the electron data are used here as the
preshower energy spectra have almost gaussian shapes.
 A polynomial fit is made to the above values of readout resolution for
      extrapolation to intermediate values of the  cluster ADC.
For cluster ADC values less than 3 MIPs, the shape of the energy spectrum
is close to a Landau distribution. For these cases
the GEANT produced MIP distribution is numerically convoluted with a
gaussian function.  The resulting distribution is matched 
with the MIP spectrum from the test beam data by adjusting the parameters 
of the gaussian function. 
 The final parameters of the gaussian function 
 are taken to represent the readout resolution.

\subsubsection{Procedure for generating simulated data}

	One of the important goals of the test beam studies is to learn
in detail about the detector and the readout cameras so that the GEANT
results in terms of energy deposition can be converted to the pad ADC
level. This data is then very similar to the actual experimental data on
lead ion interactions and can be treated in identical manner to study
specific physics issues. In particular the efficiency of photon counting,
fraction of contaminants and associated statistical and systematic errors
are all
computed by a  detailed comparison of simulated data and the original
particle distribution obtained 
from an event generator.

The steps involved in generating the simulated data from the event
generator (or from single particle inputs, as the case may be)  are
shown in
the flowchart of Fig. \ref{pmdsimulation}.
The package GWA98 implements detailed geometries of various detector
components in the WA98 experiment within the frame work of GEANT
simulation. For the PMD, the support structure with rigid frames, angular
bends and the stand were all included. These are  described in Ref.
\cite{thesis}.

 Using the MeV-ADC calibration relation
      the energy deposited in the scintillator pad in simulation
      is converted to ADC channels, equivalent to experimental data. After 
      this the ADC value
      is convoluted with a gaussian function of proper width taken from
the readout resolution curve. If the energy deposition is less than 3
      MIPs, a Landau distribution is used for convolution.
      The ADC value is then redistributed to neighbouring pads
according to the information on the light level, probability of the 
number of pads fired  and the fractional light leakage going into those 
pads as described in detail in Sec. 5.1.3.   At this stage 
      the simulated data looks very similar to the experimental data.

\subsubsection{Results of comparison}

The  MIP spectrum, obtained by simulating the response of 50 GeV $\pi^-$
incident on the PMD after folding all the effects, is superimposed on
Fig. \ref{mipspectra} for comparison with the test beam results. Both the
shape and the mean value of $\sim$ 38 ADC channels compare very well with
the test data.

The simulated preshower energy spectra corresponding to the test beam,
after  correcting for all effects as mentioned in earlier sections,
are superimposed as continuous histograms on Fig. \ref{adcspectra}
for comparison. The widths are now larger compared to
the uncorrected GEANT
results (dashed histograms). The agreement  with the test data is very
good. 

 The average number of pads fired  and the mean cluster
ADC, after applying above corrections,  are compared with
the test beam data in Fig.
 \ref{padsimdata}.
In all the cases agreement between the test beam  and the simulated data
is
excellent. These comparisons reflect the success of the methodology
adopted to generate the simulated data.

\subsection{ MIP in lead beam data}

As all the pads of the PMD could not be exposed to the test beam, it is
desired to have an alternative method to investigate the response of MIPs
in individual
pads. This is important for comparing relative pad gains and normalization
of camera gains.
   The following procedure  is adopted to obtain  the MIP
signal in the PMD from the experimental data on Pb + Pb collisions, where 
       most of the produced  
      charged hadrons  appear as MIPs.

A MIP  cluster  in test data satisfies the criteria : 

{\noindent number of pads in the cluster $\leq$ 4, } \\
fraction of total signal in the hit pad $\geq$ 0.7, \\
total cluster ADC $<$ 200.

Isolated clusters satisfying
      the above  criteria in the simulated  Pb + Pb collision data 
have been selected as
MIP signals. It is found out from the simulation study that this
      method  works quite well to identify MIPs even in the very high
      particle density environment of the lead ion experiment. The mean
and
the shape of the distribution agree  well with test data.
	The admixture of photons in the MIP sample of  simulated
lead ion data is less than
20\%. This suggests that  the algorithm for MIP
selection is satisfactory. The MIP mean is
found to be $\sim 38$ channels, in close agreement with the value
obtained from
the test beam data.

      The same prescription is also used in the experimental lead beam
data
to find out 
      the MIP signal. Fig. \ref{pbpb_mip} shows the MIP spectrum
obtained
      from experimental  and  simulated data of Pb + 
      Pb collisions at $158\cdot A$GeV. Filled circles  are
experimental data points
and  the continuous histogram represents simulated data.
MIP mean in this case  is also around 38 channels, as expected (Fig.
\ref{mipspectra}).

  \subsection{ Detector uniformity}

In the final configuration of the detector
      the response of the pad -- fibre combination further changes because
of the gain variations
      over the CCD surface and distortion in imaging leading to
      non-uniform division of pixels among fibres. This variation in  
pad-to-pad
      gain needs to be investigated.

           The gains of individual pads in the assembled detector are 
       studied in a manner
similar to that in the case of  WA93 PMD \cite{wa93nim}.
   The    MIP signal in individual pads is obtained
      from lead ion data by using the MIP criteria discussed above. The
mean value of MIP response for  
each pad is obtained by fitting the MIP ADC spectrum to a Landau distribution.
        By analyzing a large volume of data MIP spectra for all
the pads were obtained with reasonable statistics so that the mean values
are stable.
The MIP value for an individual pad
       is normalized to a global  mean of the MIP values of all the pads
in a box module, this normalization representing the relative pad gain.

The relative gains of all the pads studied are plotted in Fig. \ref{pad_gain}. 
      The fluctuation of pad-to-pad relative gain is approximately 10\%. 
This result is consistent with the measurement of pad responses 
during  fabrication  using the $\beta$-source. 
Only about 2.5\% of the pads are found to
have either very low or very high gains, lying outside the width of the
main gaussian.
These arise because 
of several possible imperfections in the detector at the fabrication and
assembly stages. 
As the fraction of such pads is
not significantly large, no extra care has been taken for their
treatment.

\section{Parametrization for the Entire PMD}

\subsection{Extension of the test results} 

The parametrizations discussed in the previous section were derived for 
 only one box module which was
tested using $e^-$ and $\pi^-$ beams at various energies.
For the simulation of the entire PMD it is thus necessary to extend the
above parametrizations in terms of suitable normalizations so that the
results are applicable for all the box
modules.
Light leakage has been extensively studied for all the box modules using
fiducial data as described in Sec. 5.1. 
The procedures for extending
the parametrization for MeV-ADC calibration and readout resolution to the
entire detector are described below.

\subsubsection{MeV--ADC relation}

In the simplest scheme,
the  MeV--ADC calibration relation can be extended to other cameras by
normalizing the polynomial obtained in Fig. \ref{mev-adc} with the
relative camera gains. Mean MIP responses in different cameras, obtained
using data on lead-lead collisions, can be used for estimating the 
relative camera gains. 

However the relation at higher light intensities may not follow the
normalization provided at the MIP point. To check this  
a  procedure involving detailed intercomparison of responses in
different cameras has been carried out at various light intensities 
using the fiducial data.

The ratio $R^i _{jk}(l)$ of signals (mean ADC values) in
pairs of cameras $j,k$ have been studied for different
intensities ($l$) of input
light from the flasher system. This ratio denotes the relative camera
gain at a particular  light intensity.
The light intensity in the fiducial event is determined by the signal in a
 camera ($i$) different from the pair under consideration. The value of
$l$ ranges from MIP level to 2500 ADC
channels. For the present study $l$ has been divided into five ranges as
shown in Fig. \ref{spatial}.

The relative gain at MIP
level is given by the ratio of MIP signals in cameras $j$ and $k$. 
If the relative gains at different light intensities are
comparable, $R^i _{jk}(l) \sim R^i _{jk}(mip)$. 
Thus one should expect that the average of the relative difference 
[$( R^i_{jk}(l) - R^i_{jk}(mip)$)] / $R^i _{jk}(mip)$ 
should be almost zero. Fig. \ref{ratios}(a) shows the histogram of the
relative difference for a large number of combinations of $i,j,k$ studied.
This is found to have  mean close to zero and a standard deviation of
 $\sim$ 8 \%,
which is rather  small and of the same order as the uncertainty in the
MeV-ADC relation. This suggests
that the relative camera gains are not very much
dependent on light intensity. Thus the   
normalization using MIP values can be considered 
satisfactory for the present case.

The normalization relation for the other box modules ($j$) is :

ADC$_j$ = [MeV $\rightarrow$ ADC$_i$] $\times$ MIP$_j$ / MIP$_i$

{\noindent where $i$ refers to the box module used in the test beam}.

\subsubsection{Readout resolution}

Data on fiducial lighting, where blue light is injected into the pad
through the fiducial fibre, provide the effects of fluctuations in all the
steps of light transmission through the fibre and the II chain to the CCD.
Thus the width of the ADC spectrum in fiducial data
($\sigma_{fid}$) should
reflect the convolution of two major components : (a) the intrinsic
fluctuation in the intensity of
light produced by the flasher system ($\sigma_{source}$) and (b)
fluctuations due to all the
other effects of
transmission of light up to the CCD ($\sigma_{trans}$), i.e.,
$ \sigma_{fid}^2 = \sigma_{source}^2 + \sigma_{trans}^2$. 
The fluctuations in the production of light (i.e., in 
energy deposition and scintillation processes), which do not contribute
to the widths of fiducial data, should be similar in all the modules as
the same scintillator has been used for the fabrication of the entire
detector. 

As the contribution of fluctuation in input light ($\sigma_{source}$) is
essentially the same for all the cameras, the difference 
$\sigma_{fid}^2 (j) - \sigma_{fid}^2 (k)$ for any pair $j,k$ of cameras
measures the difference in fluctuation due to transmission including the
effect of camera to camera variations. Fig. \ref{ratios}(b) shows the
histogram of the
deviation in width $\sqrt{|\sigma_{fid}^2 (j) - \sigma_{fid}^2 (k)|}$. This
has a mean of only 5.6\%, suggesting that the parameters of 
readout resolution are almost independent of camera gains and can be 
taken to be identical for all the cameras.

\subsection{Comparison of simulated and experimental data on Pb + Pb
collisions}

The usefulness of the extended parametrization for the entire PMD can be
tested by
comparing the results of a full simulation using the VENUS event generator
and
the experimental data on lead ion collisions. It is
well understood that particle production in lead ion collisions need not
be fully described by VENUS in all its details, but some general features
of particle emission may not be very different. Thus the average
pad occupancy and total energy deposition in simulated and experimental 
data can still be compared.

 Fig. \ref{leadresult} shows (a) the number of pads fired and (b) the
total ADC in
 box modules of different pad sizes for the experimental and
simulated data on Pb + Pb collisions. For a given pad size the values in
the figure denote the average over the number of such box modules used.
The agreement between the experimental and simulated data is within 10\%
for  both the number of pads fired and the total ADC.
This is very encouraging particularly in view of the
fact that neither any parameter of the event generator  was tuned nor the
raw experimental data was treated in any manner except noise filtering.
This suggests that the procedure for generating the simulated data and
the extended parametrization are quite satisfactory.

\section{Photon Counting}

Using the experimental data, either recorded in mode 3 (fibre or pad
coordinates) or converted to mode 3 using pixel-to-fibre maps, 
photon counting is done by first clustering the pad hits and then applying
a suitable threshold to reject hadrons.  The clustering algorithm
described in Ref. \cite{wa93nim} has been used here with some
modifications. Because of the use of pads of different sizes, the pad
matrix over the entire detector cannot be represented by a single
two-dimensional matrix to allow processing of all the pad hits in one
pass. Hence the hits are
clustered separately for each box module. 
Some tracks may produce more than one cluster either because of
upstream conversion or because of splitting at the
boundaries of the box modules. For photon tracks, the cluster with the
higher signal is
assigned a photon identity and the other one is treated as a contaminant. 

Although most of the hadrons behave like MIPs, 
a small fraction undergoing interaction in lead converter produces
signals which are similar to those of photons. Therefore all the hadrons
are not rejected by applying a threshold.
 The exact particle identification of each hit cannot be
ascertained in the experimental data. Hence the number of clusters
remaining
above the hadron rejection threshold is termed as $\gamma$-like. This
contains majority of photons and some contaminants
which reduce the purity of the photon sample.

We define the following two variables :

$\epsilon_\gamma = N^{\gamma,th} _{cls} / N_\gamma ^{inc}$ 

$f_p = N^{\gamma,th} _{cls} / N_{\gamma-like}$ 

{\noindent where $\epsilon_\gamma$ is  the photon counting
efficiency and $f_p$ is the fractional purity of the photon sample.
$N_\gamma ^{inc}$ is the number of incident VENUS photons on the PMD,  
 $N^{\gamma,th} _{cls}$
is the number of
photon  clusters above the threshold and $N_{\gamma-like}$ is the total
number of clusters above the threshold.}

An estimate of the optimum value of hadron rejection threshold and the
achievable purity of the photon sample along with the estimates of photon
counting efficiency is made by a detailed study of the simulated data,
which are generated using
VENUS event generator and GEANT simulation  following the scheme of the
flowchart
of Fig. \ref{pmdsimulation}.  No lower threshold on the energy spectrum of
photons is applied.

By adjusting the  discrimination threshold it is  possible to
obtain a reasonably pure sample
of photons, although some contaminants always  remain.
Fig. \ref{gama_eff} shows the photon counting efficiency and the purity
as a
function of hadron rejection threshold (in MIP units) for two different
centralities of Pb$+$Pb collisions. The centrality of the
reaction is defined by the transverse energy ($E_T$) obtained from the
mid-rapidity calorimeter  in the WA98 experiment \cite{wa98}.
Central events span the region
$E_T\ge 330$ GeV corresponding to the top 5\% of minimum bias cross section
and peripheral events correspond to the region $40$ GeV $\le E_T\le 100$ GeV.
In both cases the photon counting efficiency decreases with
increasing threshold. The purity improves significantly with
increasing threshold only up to $\sim$ 3 MIPs and then
rather slowly at higher thresholds.

The estimated photon counting 
efficiency is dependent on both the detector hardware and the clustering
software.  A photon is labeled as ``converted" on depositing a minimum 
energy equivalent to 0.2 MIP. The mean conversion probability   
for VENUS photons within the
PMD acceptance in the case of
Pb + Pb collisions is found to be 95\%. This gives an upper limit to the
photon counting
efficiency when no hadron rejection threshold is applied,
i.e., when all the ``converted'' photons can be counted. Photon counting
efficiency close to this value (~93\%) is achieved only for peripheral events
with no  threshold as shown in Fig. \ref{gama_eff}.
The maximum value for central events  is about 84\%. 
The decrease in photon counting efficiency for central events
arises because of loss of clusters due to overlap in the higher
multiplicity environment of central collision events.

For all practical purposes a 3 MIP threshold appears as an optimum choice
for hadron rejection leading to reasonable values for both the photon counting
efficiency and the purity. With this, the photon counting efficiencies for
central and peripheral cases are 68\% and 73\%, respectively.
The purity of the photon sample in the two
cases are 65\% and 54\%, respectively. 

The efficiency values are comparable to or better than that of the WA93 PMD.
The PMD is therefore quite suitable to handle the increased particle
density.
The purity of the photon sample is somewhat lower in the present case.
The purity depends
on the relative population of photons and hadrons as given
by the event generator. The ratio
$N_\gamma/N_{charge}$, on the PMD acceptance for central collisions
is found to be only $\sim 0.7$
in the present case compared to $\sim 0.8$ for the WA93 case.
This difference partly explains the lower purity
in the present case. A detailed investigation of the centrality and
pseudorapidity dependence
of photon counting efficiency and purity will be
presented in a subsequent publication.

For the experimental data one is able to determine only $N_{\gamma-like}$,
the number of clusters above hadron rejection threshold. Using the
estimated values of $\epsilon_\gamma$ and $f_p$, one can estimate the
number ($N_\gamma ^{est}$) of photons incident on the detector in the
event using
the relation :

$N_\gamma ^{est} = N_{\gamma-like} \cdot f_p / \epsilon_\gamma$

Statistical error on $N_\gamma ^{est}$ thus deduced is governed mainly by
the
nature of counting statistics. In the present case the values obtained 
for the statistical error are
4.6\% for central collision  events, where the average number of
incident VENUS
photons, $<N_\gamma ^{inc}>$, is 428, and 11\% for peripheral collision
events
having $<N_\gamma ^{inc}> \sim$ 116. Details regarding the systematic
errors will be discussed in a future communication.

Recently it has been reported \cite{nn_nim} that using neural network
techniques and
considering the full information of preshower profile, much better
discrimination between photons and hadrons can be achieved. 
Thus one can improve the purity of the photon sample without sacrificing
the photon counting efficiency. In this method each cluster is assigned an
output value optimized by the network algorithm and one can select photons
and hadrons by applying suitable threshold on the network output spectrum. 
Full potential
of this method and its application to actual experimental data is still
under investigation and will be reported in subsequent publications.

\section{Summary}

The preshower PMD for the WA98 experiment has been fabricated and tested.
The detector consists of 3 X$_0$ thick lead converter plates and a matrix
of plastic scintillator pads placed behind the converter in 28 box
modules. Square pads of four sizes varying from 15 mm to 25 mm have been
employed. Scintillation light is transmitted to the readout device using a
short WLS fibre spliced with a long  EMA-coated  clear fibre to improve
light
transmission. Each box module consists of a matrix of 38 $\times$ 50 pads
and is read
out using one II + CCD camera system obtained from the
UA2 experiment. The detector extends to 21 m$^2$ in area and covers the
pseudorapidity region 2.5 $\le \eta \le$ 4.2 of which the region 
3.2 $\le \eta \le$
4.0 has full
azimuthal coverage.

The imaging of the fibres on the CCD surface  has been greatly improved by
several modifications in the techniques of detector fabrication and
assembly. This has  resulted in fewer number of pads fired for a
preshower
of a given electron energy as compared to that in the case of WA93 PMD. 
In addition, the granularity has been
optimized so that the pad occupancy and the multiple hit
probability are quite low. These efforts have
improved the performance of the detector in the much higher particle
density environment of lead ion collisions. 

The detector was tested using electron and pion beams at different
energies to study the preshower behaviour and the
response of minimum ionizing particles. 
Data were also taken using fiducial lighting at several
 light intensities for all the readout units in the box modules.
The test beam and fiducial data have been used to deduce the parameters
describing the behaviour of the detector. 
A simulation framework incorporating these parameters has been developed
to generate simulated data which are very similar  to the
experimental data and can be compared directly to study the physics with
event generators. The simulated and the experimental data compare very
well for both the test beam runs and for gross features of the lead beam
runs.

The photon counting efficiency is found to be 68\% for the central and
73\% for the peripheral Pb$+$Pb collision events as given by the
VENUS event generator. The slight decrease in the efficiency for central
collision case is due to loss of clusters because of overlap.
The purity of the photon sample is about
65\% for central and 54\% for peripheral cases.

{\bf Acknowledgements}

           We gratefully acknowledge the financial support of the 
      Department of Atomic Energy, the Department of Science and
Technology
      and the University Grants Commission of the Government of India.
	Two of us (DSM and NKR) are grateful to the Council of Scientific
and
Industrial Research  for the financial support for
carrying out this work. We express our sincere gratitude to CERN
authorities for the loan of UA2 readout cameras and for extending the
servicing facilities for them. We thank C.
Engster and J. Dupont  for  help
in testing  the UA2 cameras and P. Pierre and T. Reynes 
for the engineering support  during the installation of the PMD.
We are grateful to the PPE Division of CERN  for providing  local
hospitality to the Indian team 
and to the Indo-FRG Exchange Programme  for supporting the visits of
collaborators from GSI, Germany to India. 
 We also acknowledge the
cooperation of the engineering, computer and other support staff at the
collaborating institutions (VECC, IOP, Jammu
University, Panjab University and Rajasthan University) during the 
period of design, fabrication and assembly of this detector. We express
our sincere gratitude to the CERN SPS crew for providing good beams during
test runs.


%

\clearpage

\parskip=0pt

\begin{figure}
\centerline{\psfig{figure=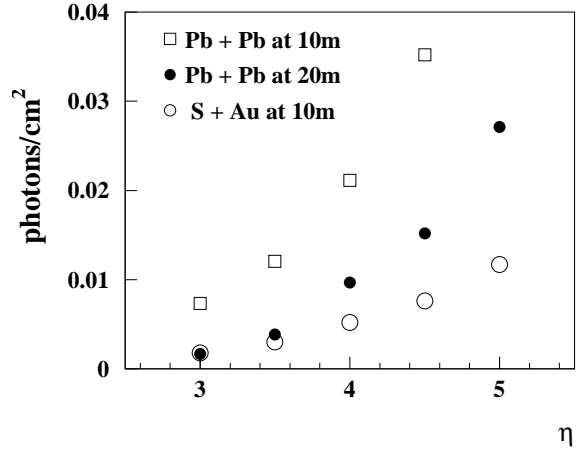,width=8cm}}
\caption{\label{density}
 Particle density (photon/cm$^2$) as a function of pseuedorapidity
($\eta$) in central lead-lead collisions as given by VENUS.
 }
\end{figure}

\begin{figure}
\centerline{\psfig{figure=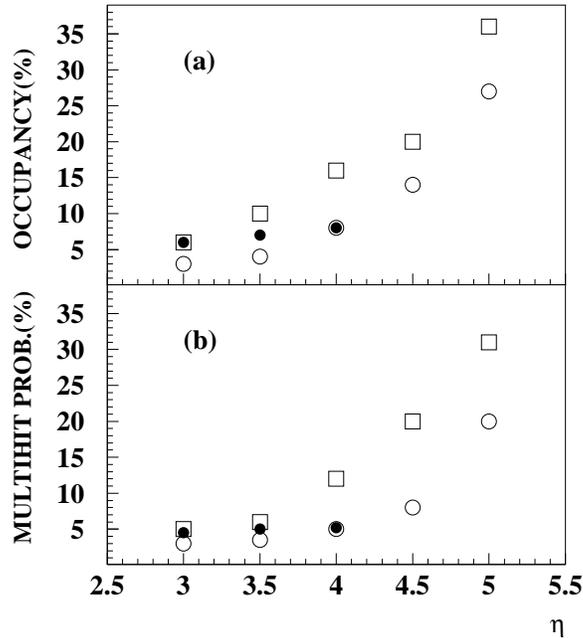,width=8cm}}
\caption{\label{multihit}
(a) Occupancy and (b) multiple hit probability for 
three different cases: 15 mm
pads (open circles), 25 mm pads (open squares) and the final configuration 
having four different pad sizes (filled circles). 
The distance from the target was 21.5 m for the final configuration and 20 m 
for the other cases.
 }
\end{figure}

\clearpage

\begin{figure}
\centerline{\psfig{figure=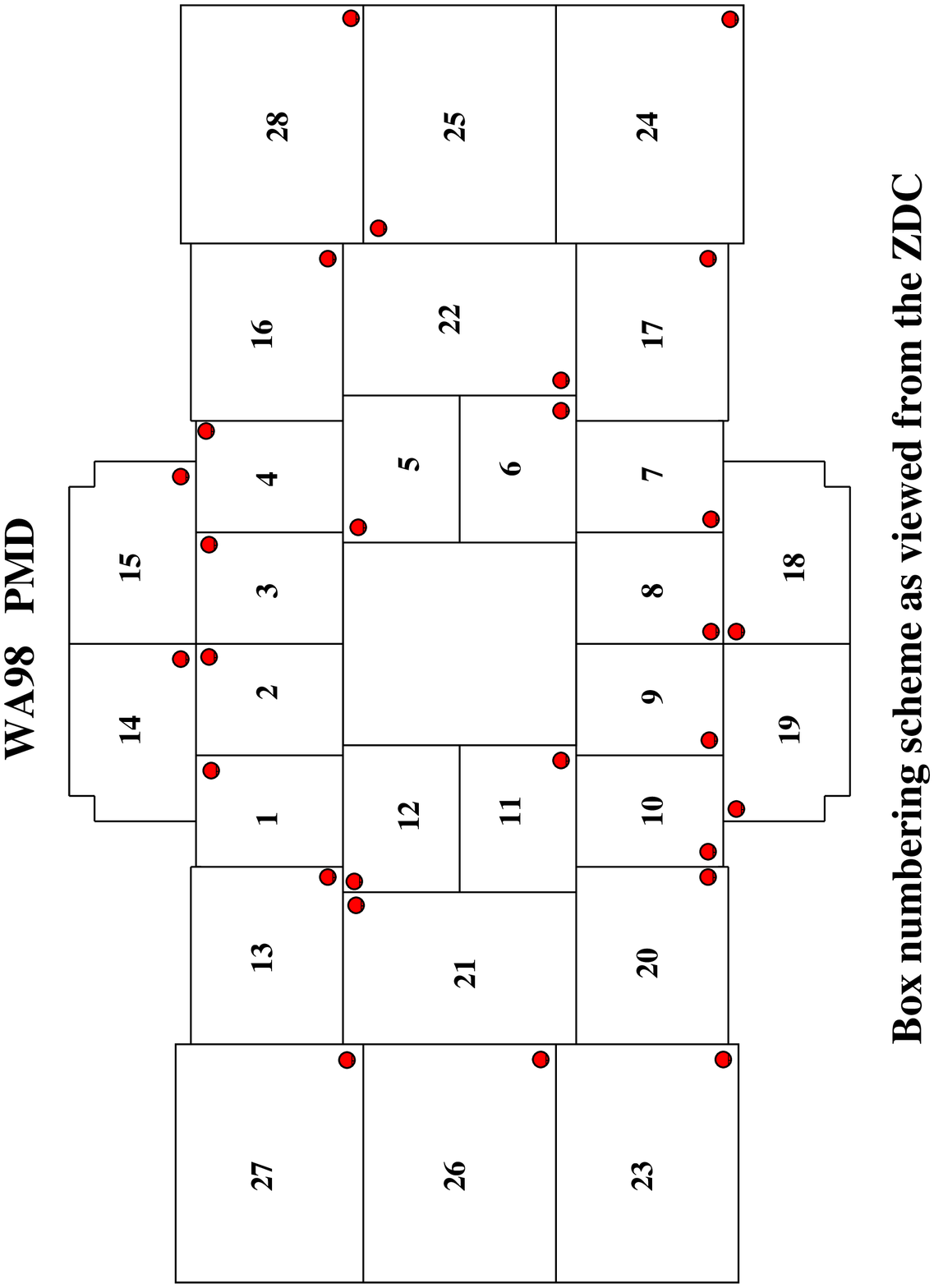,width=12cm,angle=-90}}
\caption{\label{layout}
  Layout of the detector box modules. The black dot in each box indicates
the corner where  the pad with the co-ordinate (01,01) is located. Box
modules
numbering 1-12 have 15 mm pads, 13-20 have 20 mm pads, 21-22 have 23 mm
pads and the rest have 25 mm pads.
 }
\end{figure}

%
\begin{figure}
\centerline{\psfig{figure=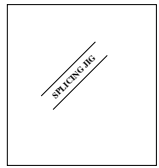,width=2cm}}
\caption{\label{murthy2}
  Schematic of the splicing jig arrangement.
 }
\end{figure}

\begin{figure}
\centerline{\psfig{figure=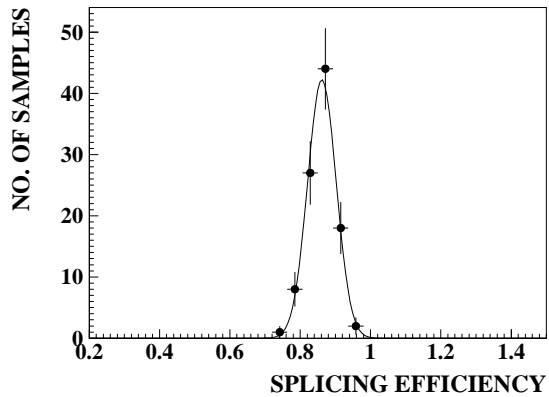,width=8cm}}
\caption{\label{trans_coef}
  Splicing efficiency ( fraction of light transmitted across the
joint) for the case of  8 cm long WLS fibre joined  with 192 cm long clear
coated fibre.
 }
\end{figure}

\begin{figure}
\centerline{\psfig{figure=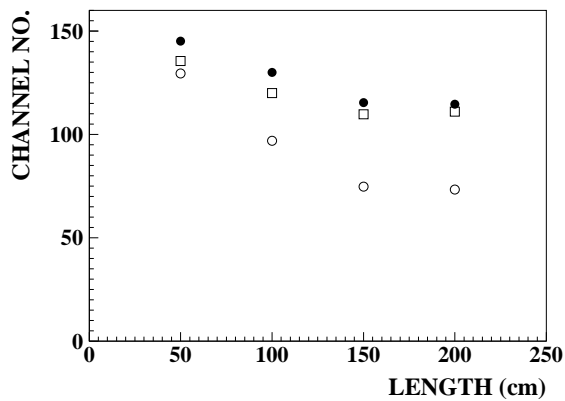,width=8cm}}
\caption{\label{op_length}
Light output of the scintillator pads (expressed as peak channel number of
the
digitized PMT signal) as a function of fibre length in
three different cases :  
open circles represent the pads with full length WLS fibre,
open squares represent the pads having 8
cm long WLS fibre piece spliced with clear coated fibre,
and filled circles represent the pads with 
WLS (8 cm) + clear fibre  and  reflector paint applied to the tip
of the WLS fibre
inside the pad. }
\end{figure}

\begin{figure}
\centerline{\psfig{figure=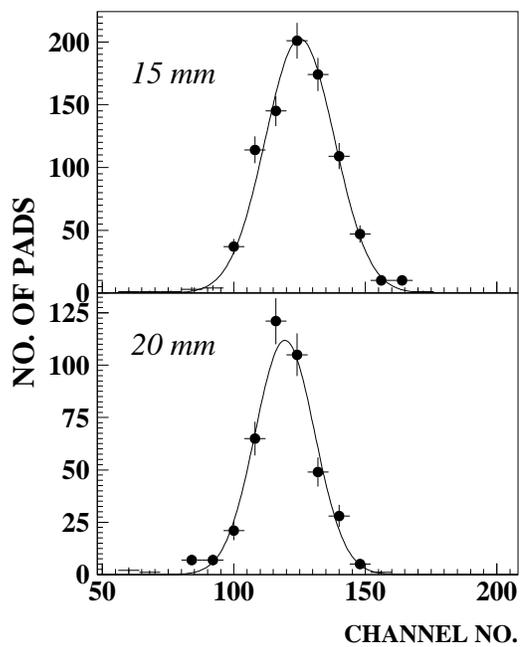,width=8cm}}
\caption{\label{beta}
 Response of pads to $^{106}$Ru $\beta$-rays. Filled circles are the data
points showing  distributions of 
peak channel numbers of the PMT signals. Continuous lines are the gaussian
fits.  Pad sizes are as indicated within the boxes. }
\end{figure}

%
%
%

\begin{figure}
\centerline{\psfig{figure=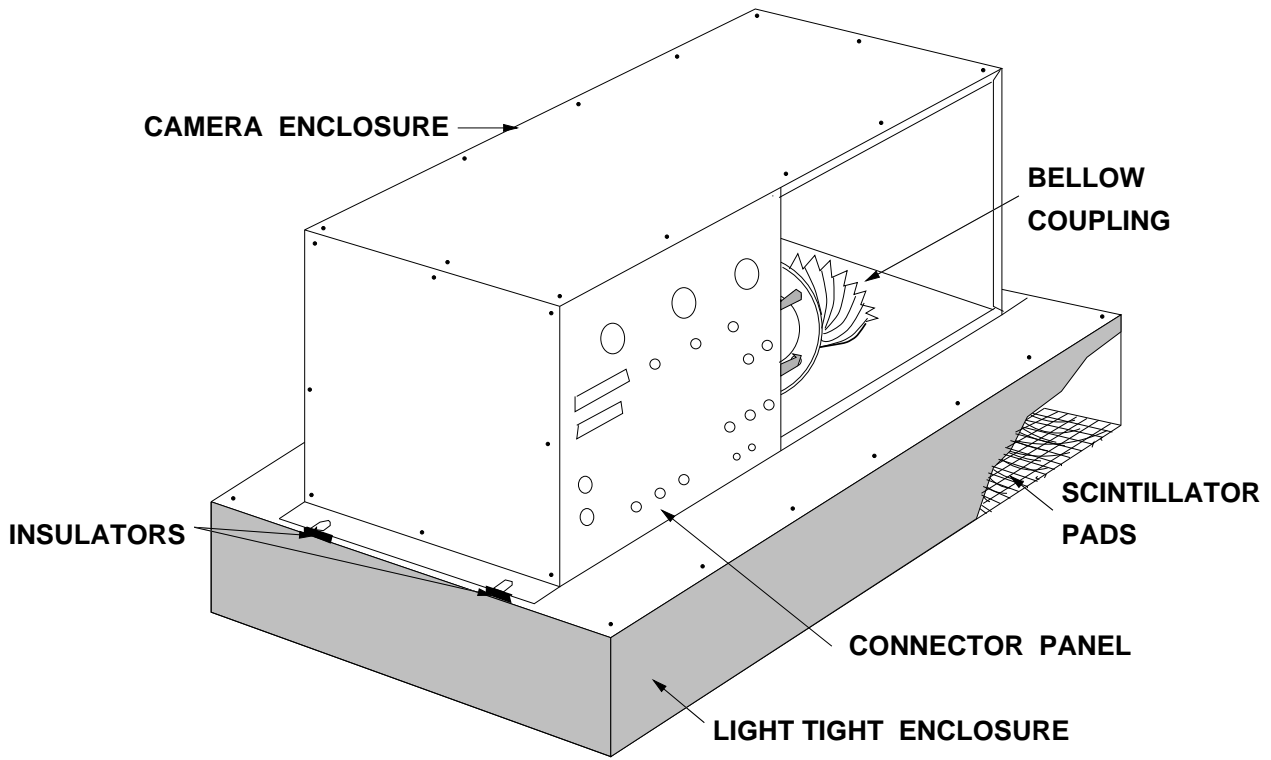,width=10cm,angle=-90}}
\caption{\label{module}
  Sketch showing a detector box module consisting of a light tight
enclosure for scintillator pads  and the camera  enclosure on
top. }
\end{figure}

\begin{figure}
\setlength{\unitlength}{1mm}
\begin{picture}(140,50)
\put(10,0){
\epsfxsize=5cm
\epsfysize=3cm
\epsfbox{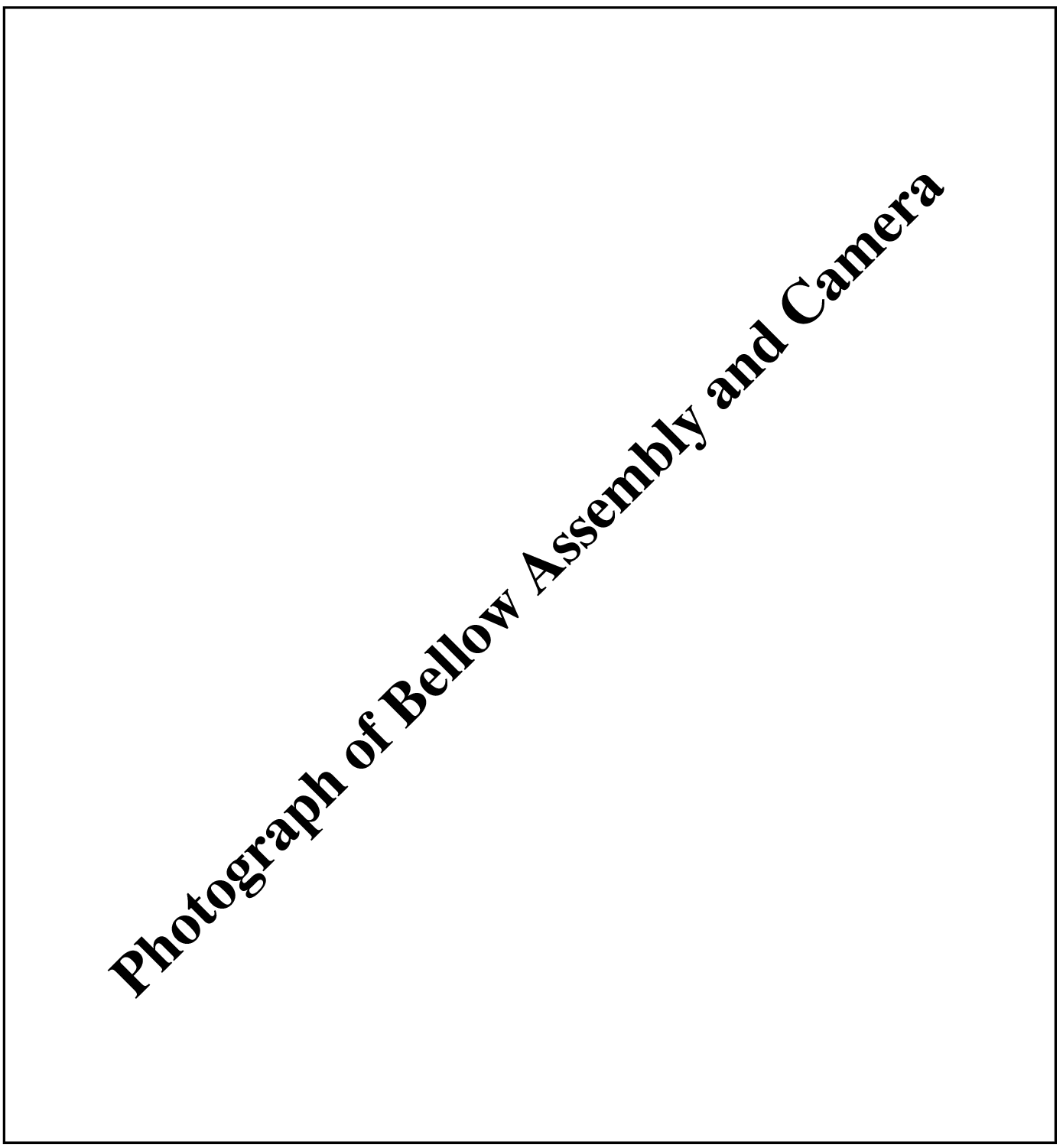}
}
\put(80,0){
\epsfxsize=5cm
\epsfysize=3cm
\epsfbox{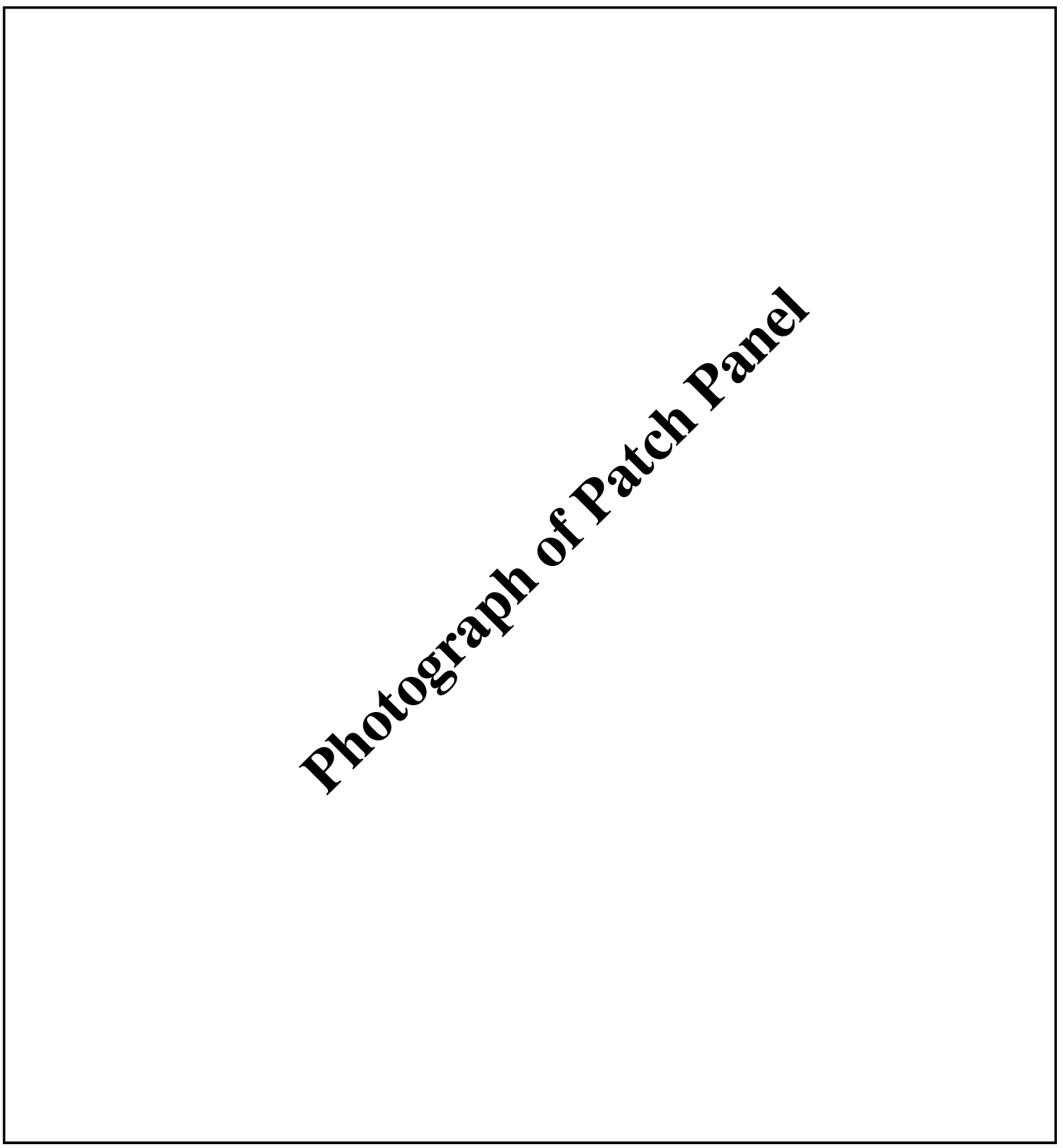}
}
\end{picture}
\caption{\label{photo1}
  Photographs showing (a) a section of the detector box module with the
rubber bellow and the II + CCD camera system coupled with the FEC plate,
and (b) the connector panel for the electrical and optical connections.
}
\end{figure}

\begin{figure}
\centerline{\psfig{figure=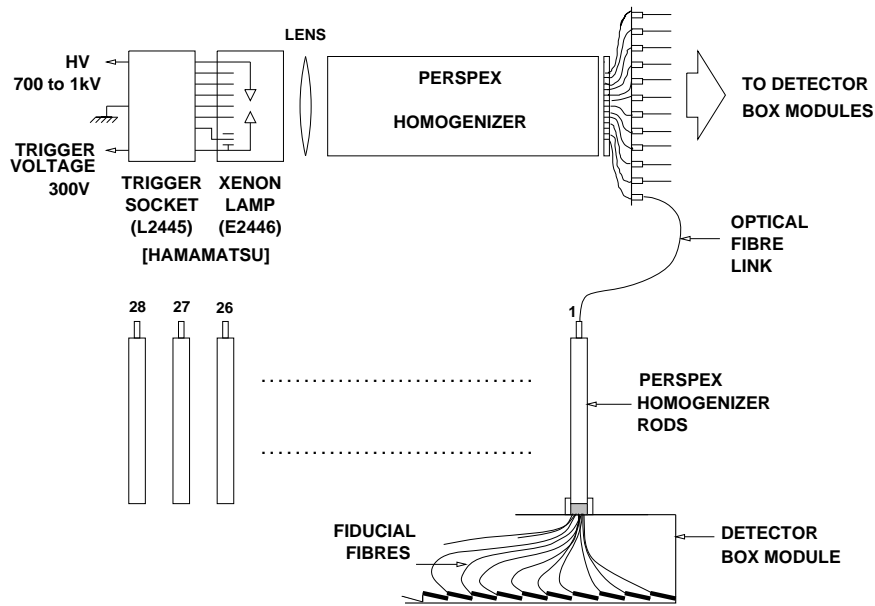,width=8cm,angle=-90}}
\caption{\label{xenon}
  Schematics of Xenon flasher and the optical distribution system for
injecting externally triggered light into the fiducial pads.}
\end{figure}


\begin{figure}
\setlength{\unitlength}{1mm}
\begin{picture}(140,20)
\put(10,0){
\epsfxsize=5cm
\epsfysize=2cm
\epsfbox{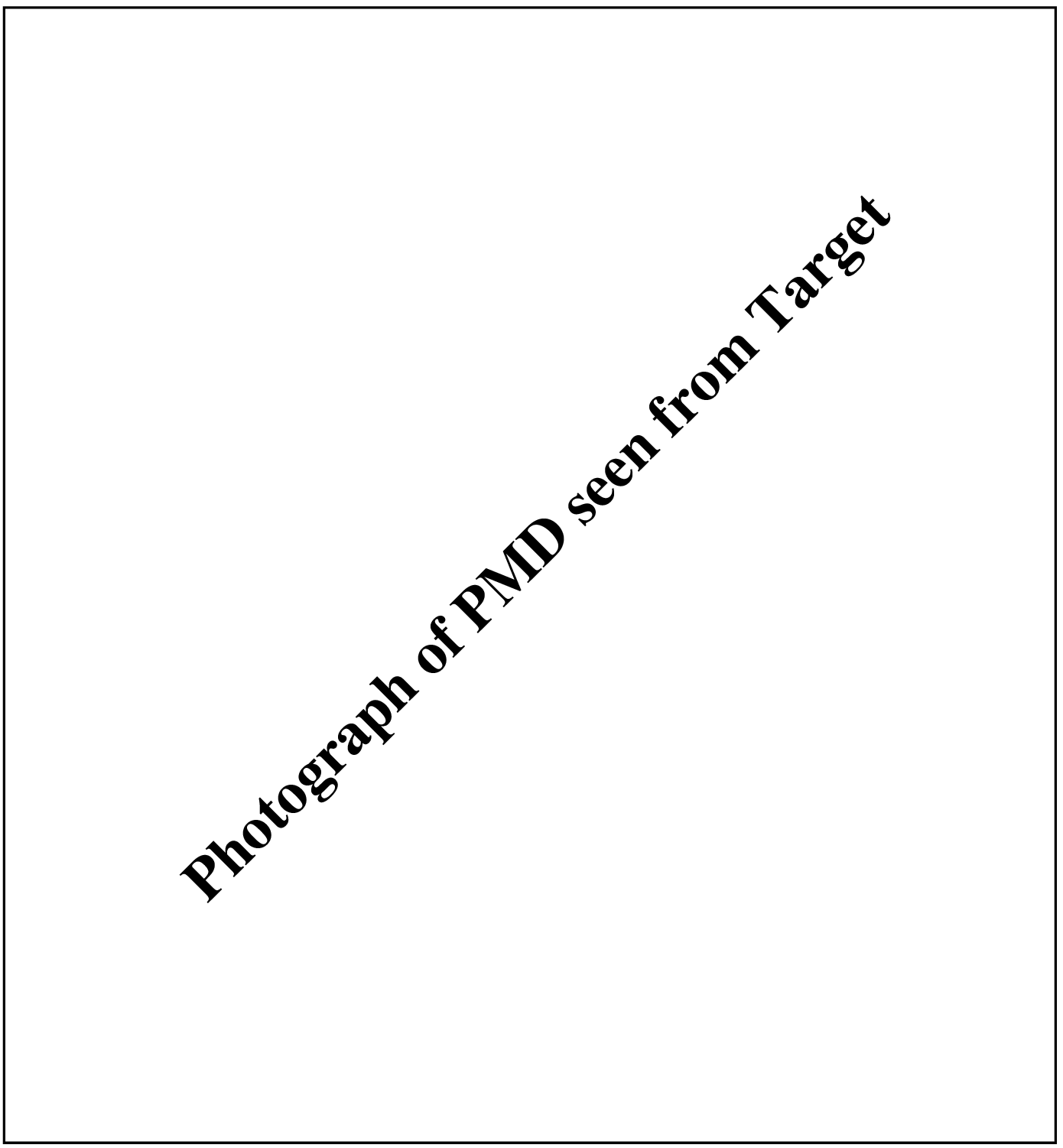}
}
\put(80,0){
\epsfxsize=5cm
\epsfysize=2cm
\epsfbox{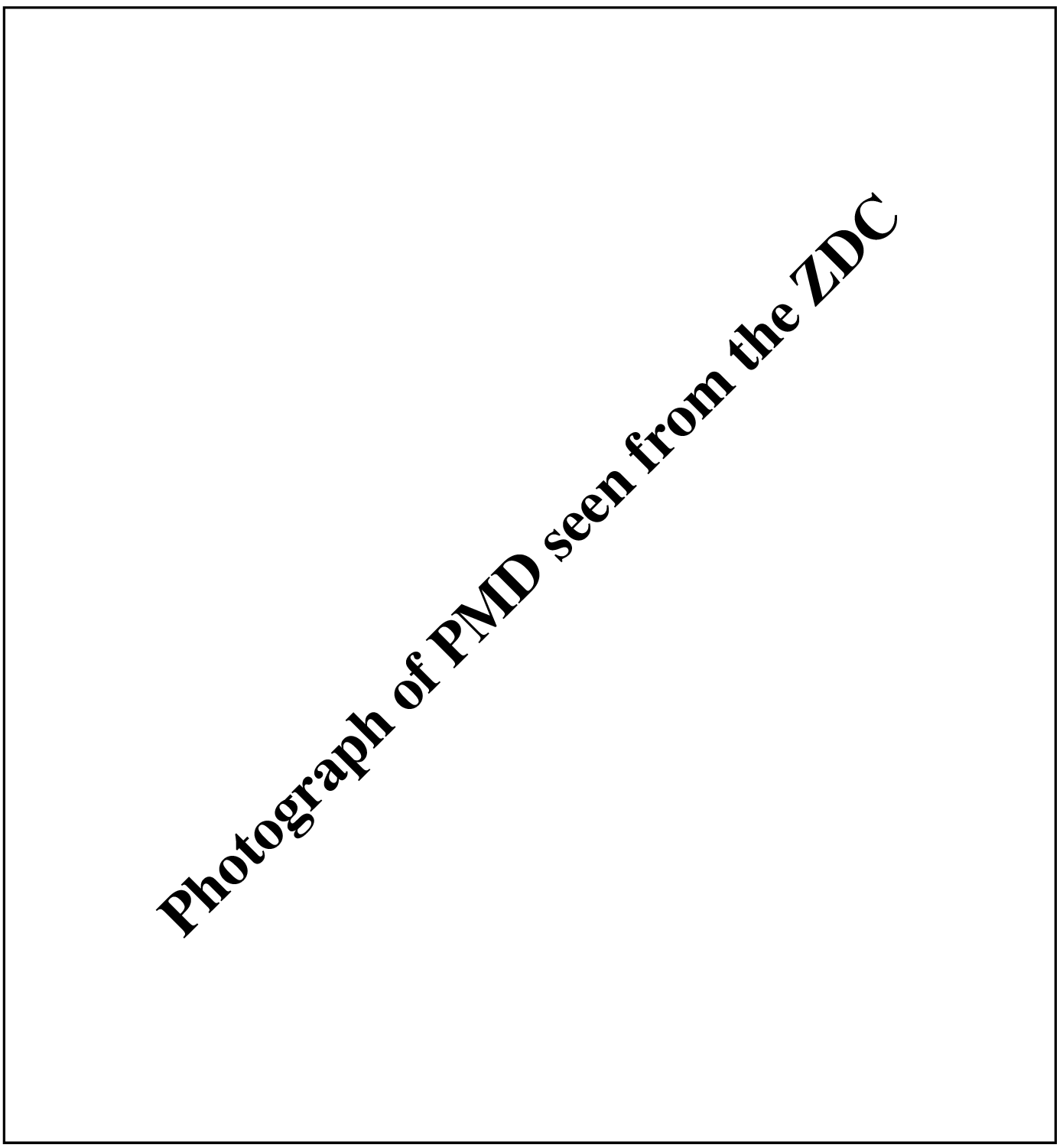}
}
\end{picture}

\caption{\label{murthy3}
   Photographs showing (a) the front side of the PMD, as viewed from the
target, with the  support structure and lead converter plates,
and (b) back side of the PMD showing the detector box modules
(orange
colour) and the camera enclosures (blue colour). }
\end{figure}

\begin{figure}
\centerline{\epsfig{figure=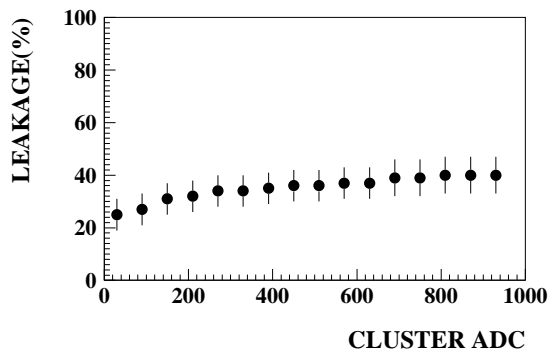,width=8cm}}
\caption{\label{adcleak}
   Percentage of light leaking outside the assigned pixels of
   a given  fibre as a function of total light (cluster ADC).
   The error bars denote the r.m.s. of the distribution of leakage.
 }
\end{figure}

\begin{figure}[h]
\centerline{\epsfig{figure=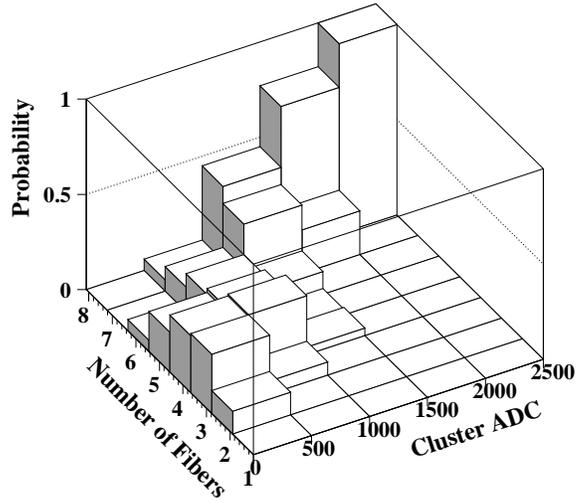,width=8cm}}
\caption{Probability distribution of the number of fibres surrounding the
fiducial one affected by
light leakage
for various light intensities as denoted by the cluster ADC.}
\label{spatial}
\end{figure}

\begin{figure}
\centerline{\epsfig{figure=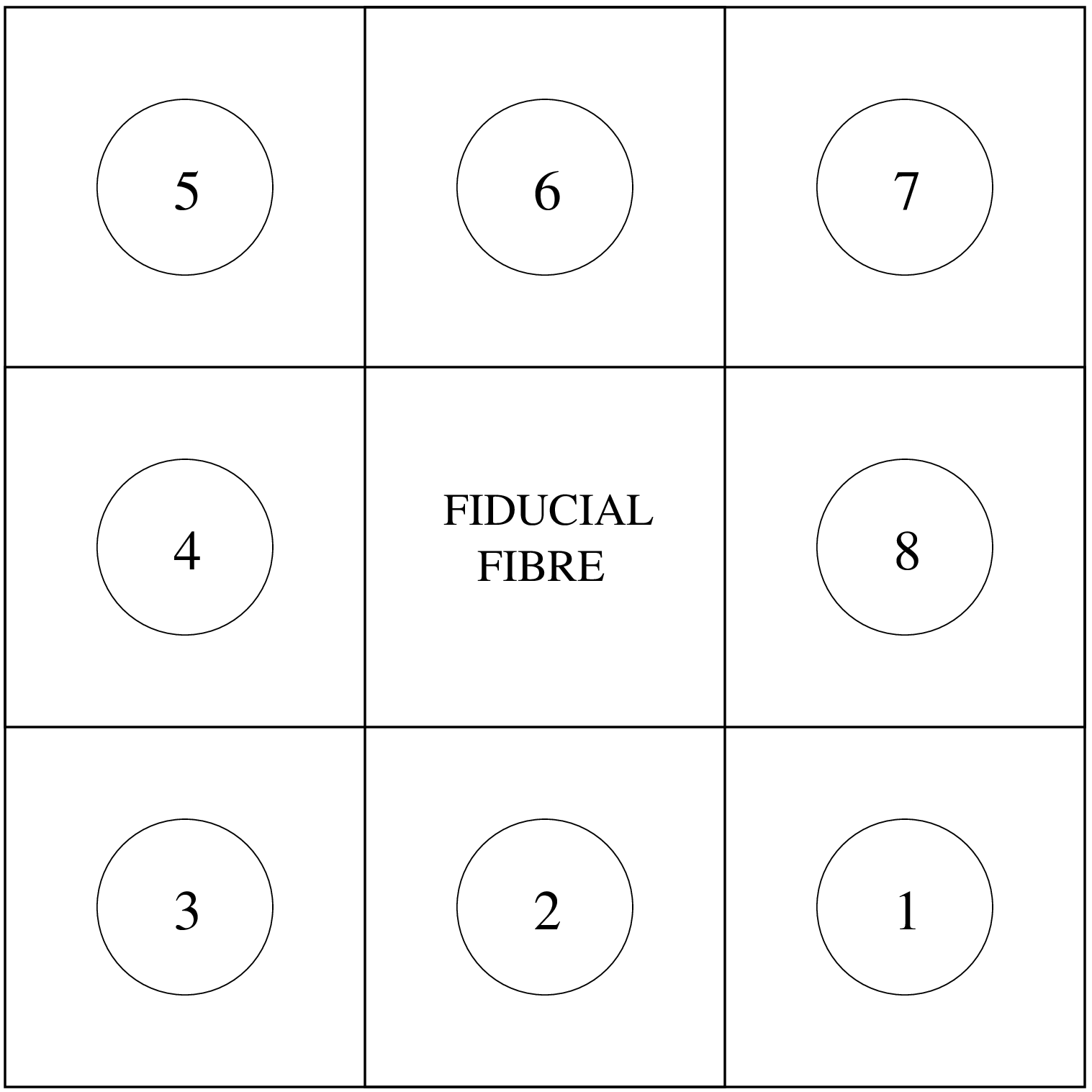,width=4cm}}
\caption{Position  of the  eight  fibres surrounding the fiducial one in a
rectangular matrix. This nomenclature is used in Table II, see text.}
\label{pattern}
\end{figure}

\begin{figure}
\centerline{\psfig{figure=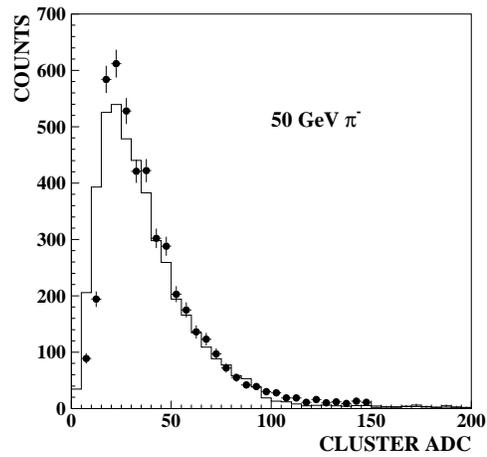,width=7cm}}
\caption{\label{mipspectra}
  ADC spectrum of 50 GeV $\pi^-$ passing through a PMD box module
without converter. Filled circles denote the test beam data.  The
continuous histogram
  represents simulation results after using MeV-ADC conversion and
accounting for readout resolution and leakage as discussed in the text.
 }
\end{figure}

\begin{figure}
\begin{center}
\epsfig{figure=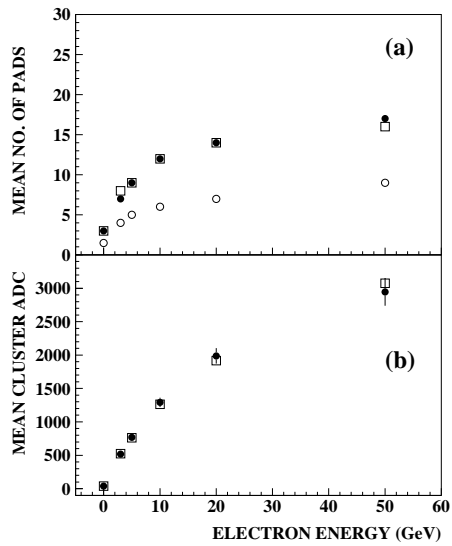,height=8cm}
\caption{\label{padsimdata}
  Preshower characteristics : (a) Mean number of pads fired 
  as a function of incident electron energy.
   Filled circles denote the test beam data, 
   open circles correspond to the GEANT simulation before 
   light leakage to the surrounding fibres, and open 
   squares represent the simulation results after introducing 
   light leakage to the surrounding fibres.
   (b) Mean cluster ADC as a function of incident electron energy.
   Filled circles denote the test beam data and open
   squares represent the simulation results after folding all
   the effects.
   The results shown are for 15 mm pad sizes.
   The point at zero energy represents the MIP case.
 }
\end{center}
\end{figure}

\begin{figure}
\begin{center}
\epsfig{figure=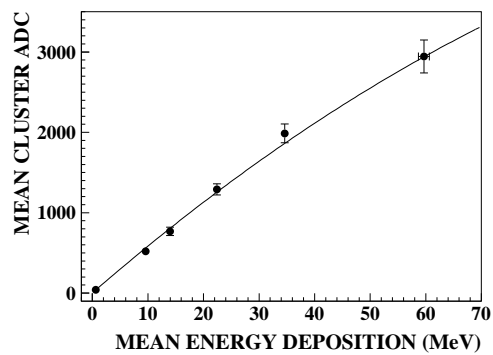,width=7cm}
\caption{\label{mev-adc}
   Mean cluster ADC of the test beam data vs. the mean energy 
   deposition from GEANT simulation. The lowest point
   corresponds to 50 GeV $\pi^-$ (a MIP) and other points correspond 
   to electrons at incident energies of 3, 5, 10, 20 and 50 GeV.
The continuous curve is a polynomial fit to the data and represents the
MeV-ADC calibration relation.
 }
\end{center}
\end{figure}

\begin{figure}
\begin{center}
\epsfig{figure=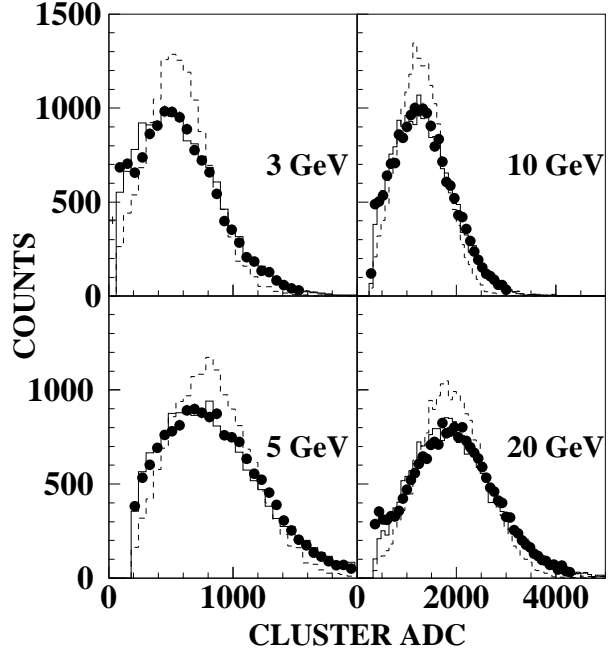,width=9cm}
\caption{\label{adcspectra}
   The preshower energy spectra for electrons at different energies as
indicated. The filled circles denote 
   the test beam data.  The dashed  histograms represent  the simulated
spectra 
   before readout resolution and the continuous histograms represent the 
   simulated spectra after folding the effects of readout resolution.
 }
\end{center}
\end{figure}

\begin{figure}
\begin{center}
\epsfig{figure=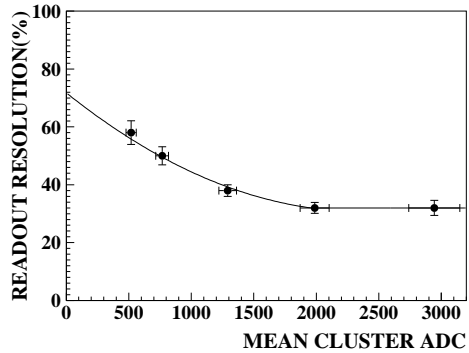,width=7cm}
\caption{\label{readoutres}
   Readout resolution as a function of mean cluster ADC.
   The continuous line represents a second order polynomial fit to the
data points  
   up to 2000 ADC and a straight line fit beyond 2000 ADC.
 }
\end{center}
\end{figure}

\clearpage
\begin{figure}
\begin{center}
\epsfig{figure=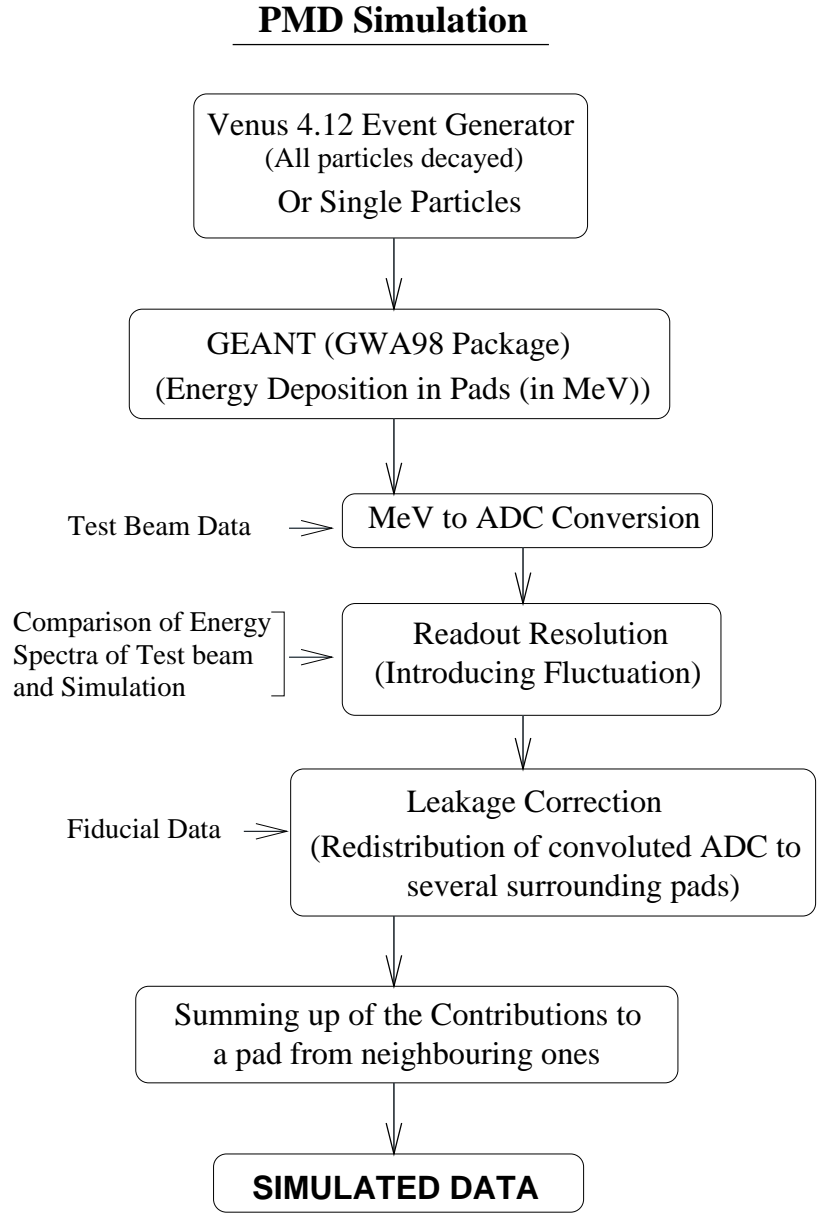,height=16cm}
\caption{\label{pmdsimulation}
   Flow chart showing various steps in generating the simulated data.
 }
\end{center}
\end{figure}


\begin{figure}

\centerline{\psfig{figure=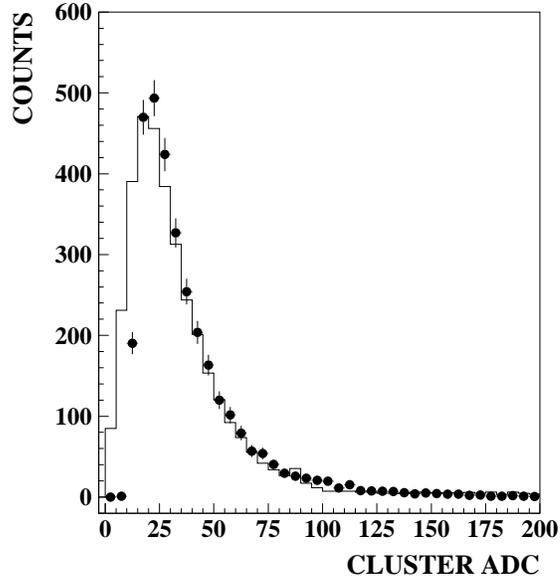,width=8cm}}
\caption{\label{pbpb_mip}
   The MIP spectra derived from the experimental data on lead-lead
collisions (filled circles) and from the simulated data (continuous
histogram). }
\end{figure}

\begin{figure}
\begin{center}
\epsfig{figure=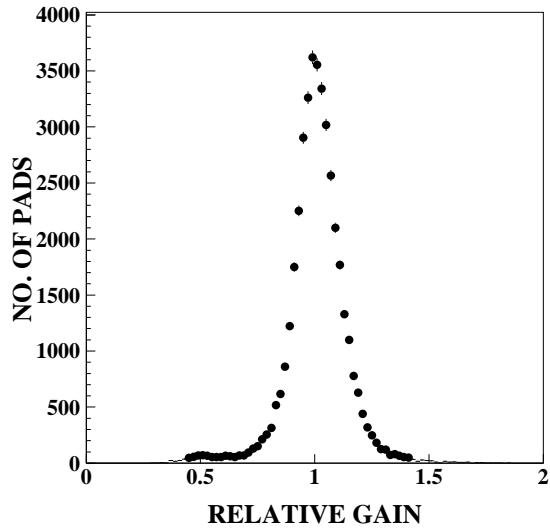,width=8cm}
\caption{\label{pad_gain}
    Distribution of relative pad gains for the entire detector. The width
($\sigma$) of the main gaussian is 10\%.  }
\end{center}
\end{figure}

\begin{figure}
\centerline{\epsfig{figure=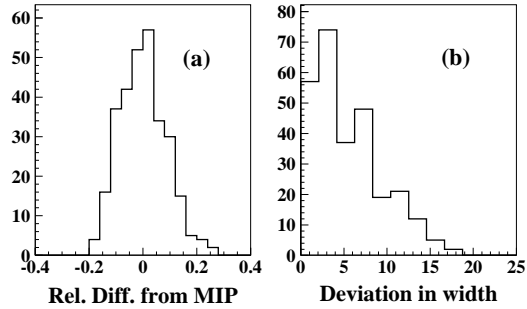,width=8cm}}
\caption{Histograms of (a) the relative difference  of the
 ratio of signals in  pairs of cameras at different light intensities to
that at MIP light level as defined in the text and (b) the deviation in widths
for pairs of cameras at different light intensities.}
\label{ratios}
\end{figure}

\begin{figure}
\centerline{\epsfig{figure=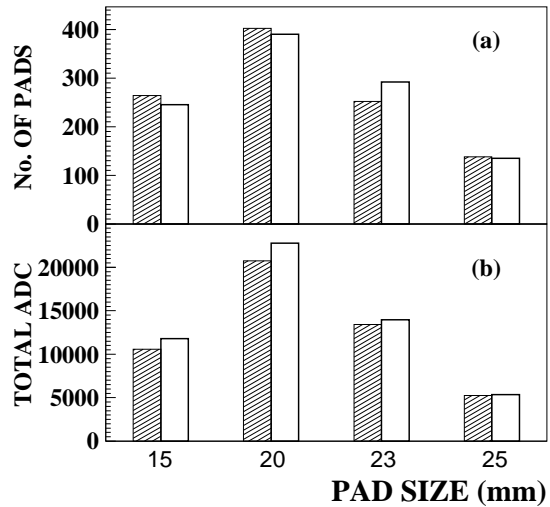,width=8cm}}
\caption{Comparison of experimental and simulated data on Pb + Pb
collisions for box modules of different
pad sizes : (a) number of pads fired and (b)
total ADC in an
event. The shaded bars denote the experimental data.}
\label{leadresult}
\end{figure}

\begin{figure}
\begin{center}
\epsfig{figure=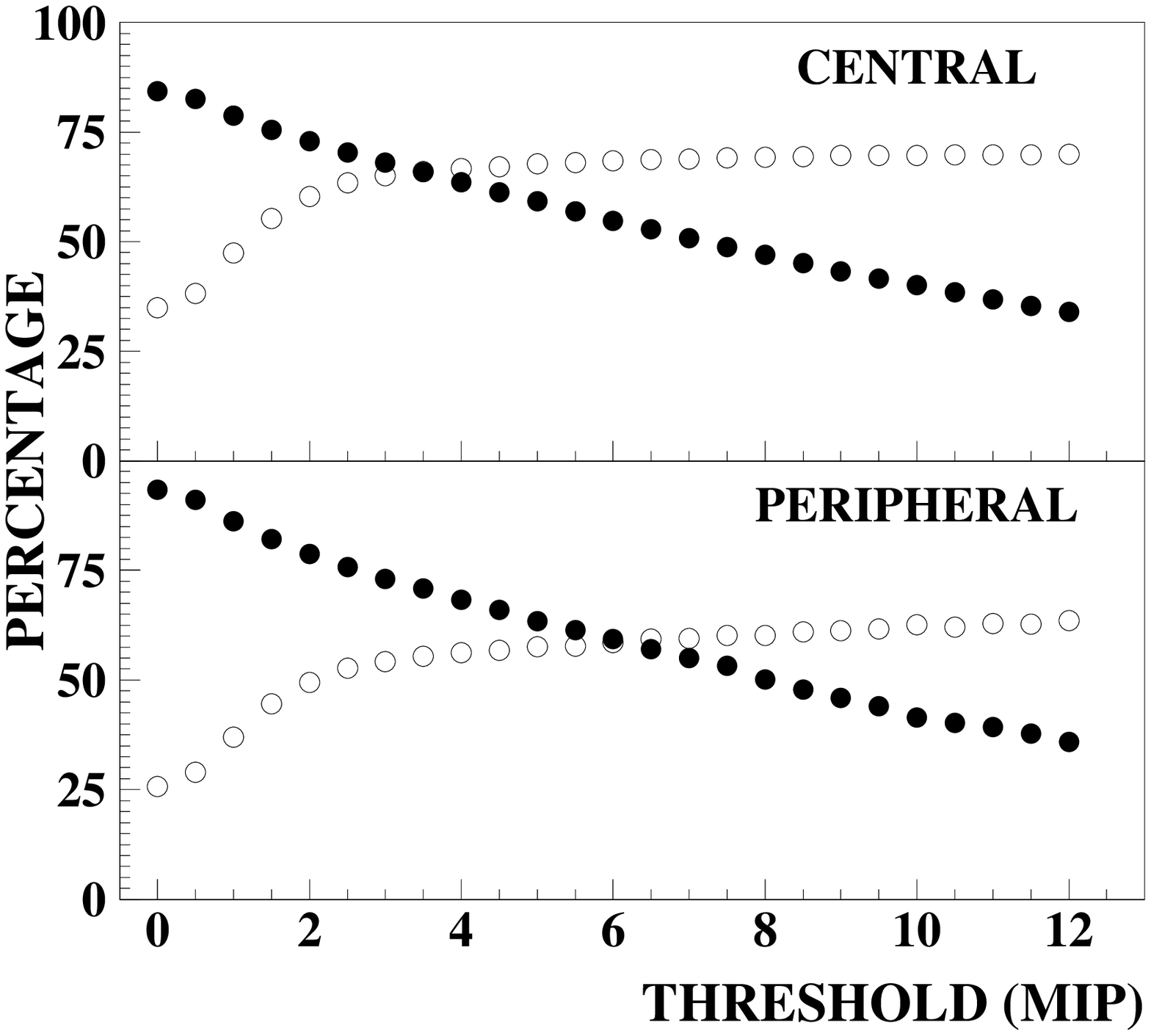,width=8cm}
\caption{\label{gama_eff}
   Variation of the photon counting efficiency (filled circles)
   and the fractional purity (open circles) with hadron rejection
threshold (in
   MIP units). Top part shows the case of central 
   events and the bottom part is for the peripheral events.
 }
\end{center}
\end{figure}


\begin{thebibliography}{99}


\bibitem{dks} E.L. Feinberg, Nuovo Cimento {\bf A34} (1976) 391; \\
              E.V. Shuryak, Phys. Lett. {\bf B79} (1978) 135; \\
              D.K. Srivastava and B. Sinha, Phys. Rev. Lett. {\bf 73}
(1994)
2421.
\bibitem{anselm}    A.A. Anselm and M.G. Ruskin, Phys. Lett. {\bf B266} 
      (1991) 482; \\
                    J.D. Bjorken, K.L. Kowalski, C.C. Taylor, ``Baked Alaska'',
                    SLAC-PUB-6109 (1993). \\ 
  J.-P. Blaizot and A. Krzywcki, Phys. Rev. {\bf D46} (1992) 246. \\
  K. Rajagopal and F. Wilczek, Nucl. Phys. {\bf B399} (1993) 395. 
\bibitem{centauro}  C.M.G. Lates, Y. Fujimoto, and S. Hasegawa, 
Phys. Rep. {\bf 65} (1980) 151. \\
Y. Takahasi et al., (JACEE Collab.) Proc. 7$^th$ Int'l. Symp. on Very High
Energy Cosmic Ray Interactions, (1992), Ann Arbor, Michigan, ed. L. Jones.

\bibitem{saphir} H. Baumeister et al, Nucl. Instr. Meth. {\bf A292} (1990)
81.
\bibitem{wa93nim} M. M. Aggarwal et al., Nucl. Instr. and Meth. {\bf A372} (1996)
143.
\bibitem{wa98} Proposal for a Large Acceptance Hadron and Photon Spectrometer,
   H.H. Gutbrod et. al., CERN-SPSLC-91-17, CERN-P260 (1991) 87 p.
\bibitem{ua2}     R. E. Ansorge et al., Nucl. Instr. and Meth. {\bf A265} (1988) 33.
\bibitem{werner}  K. Werner, Phys. Rep. {\bf C232} (1993) 87.
\bibitem{geant}   R. Brun et al., GEANT3 user's guide, CERN/DD/EE/84-1
(1984).
\bibitem{mrdm} M. R. Dutta Majumdar et al., DAE Symp. Nucl. Phys.
(Calicut, India), {\bf
 36B} (1993) 410;  DAE Symp. Nucl. Phys. (Bhubaneswar, India), {\bf 37B}
(1994) 483; Int'l. Nucl. Phys. Symp,  (Bombay, 1995) Book of
Abstracts, I-45.
\bibitem{thesis} B.K. Nandi, Thesis, under preparation.
\bibitem{nn_nim} S. Chattopadhyay et al, preprint hep-ex/9807012,
submitted to
Nucl. Instr. Meth. 

\end{thebibliography}
\end{document}